\documentclass[letterpaper, hidelinks, final,12pt]{article}
\usepackage[reqno]{amsmath}
\usepackage[psamsfonts]{amsfonts}
\usepackage{amsthm}
\usepackage[pdftex,final,debugshow]{graphicx}
\usepackage[left=2cm,top=2cm,right=2cm]{geometry}
\usepackage{amssymb}
\usepackage{amscd}
\usepackage[onehalfspacing]{setspace}
\usepackage{amstext}
\usepackage{appendix}
\usepackage{fontsmpl}
\usepackage{indentfirst}
\usepackage[colorlinks=false,linkcolor=black,urlcolor=black]{hyperref}
\usepackage{array}
\usepackage{placeins}
\usepackage{pgfplots}
\usepackage{subcaption}
\usepackage{amsmath, amssymb}
\usepackage{array}
\usepackage{booktabs}
\usepackage{multirow}
\usepackage{tabularx}
\usepackage{tikz}
\usetikzlibrary{backgrounds, calc, fit, intersections, patterns, positioning}
\usepackage{verbatim}
\usepackage{url}

\usepackage{hyperref}

\pgfplotsset{width=10cm,compat=1.9}
\makeatletter
\def\amsbb{\use@mathgroup \M@U \symAMSb}
\makeatother

\newtheorem{timing*}{Timing}

\newtheorem{definition}{Definition}

\newtheorem{lemma}{Lemma}

\newtheorem{proposition}{Proposition}

\newcommand{\overbar}[1]{\mkern 1.5mu\overline{\mkern-1.5mu#1\mkern-1.5mu}\mkern 1.5mu}

\begin{document}

\title{Never Say Never: Optimal Exclusion and Reserve Prices with Expectations-Based
Loss-Averse Buyers\thanks{%
For helpful comments, we thank Zachary Breig, Jeff Ely, Leslie Marx and audiences at the 2021 Australasian Economic
Theory Workshop at the University of Sydney, UTS, the 2021 INFORMS
Annual Meeting in Anaheim, the $12^{\text{th}}$ Conference on Economic Design in Padova, the 2022 Asia-Pacific Industrial Organization Conference in Sydney, and the 2023 Workshop on Preferences and Bounded Rationality at the BSE Summer Forum. We are also grateful to Andrew Barr, Gladys
Berejiklian and, especially, Scott Morrison for the extended lockdowns that
vastly improved our productivity. Rosato gratefully acknowledges financial
support from the Australian Research Council (ARC) through the ARC Discovery
Early Career Researcher Award DE180100190.}}
\author{\textsc{Benjamin Balzer\thanks{%
UTS Business School, University of Technology Sydney
(Benjamin.Balzer@uts.edu.au).}} \and \textsc{Antonio Rosato\thanks{
University of\ Queensland, Universit\`{a} di Napoli Federico II and CSEF
(a.rosato@uq.edu.au).}} }
\date{%
%TCIMACRO{\TeXButton{Today}{\today}}%
%BeginExpansion
\today%
%EndExpansion
}
\maketitle

 \begin{abstract}
 	%We characterize the optimal reserve price in %first-price and second-price
 	%auctions with independent private values when %bidders are expectations-based
 	%loss averse. Under \textquotedblleft unacclimating personal
 	%equilibrium\textquotedblright\ (UPE), whereby %bidders keep their
 	%expectations fixed when choosing their bids, the %optimal public reserve
 	%price can be lower than under either risk %neutrality or risk aversion.
 	%Moreover, secret and random reserve prices raise %more revenue than public
 	%ones since, by giving every bidder a small chance %to win the auction, the
 	%seller exposes all bidders to the  ``attachment %effect''. In contrast, under
 	%\textquotedblleft choice-acclimating personal %equilibrium\textquotedblright\
 	%(CPE), whereby a bid determines both the reference %lottery and the outcome
 	%lottery, the optimal reserve price is public and it %differs across the two
 	%auction formats. Furthermore, the seller excludes %more types compared to the
 	%risk-neutral and risk-averse benchmarks.

\noindent We study reserve prices in auctions with independent private values when bidders are expectations-based loss averse. We find that the optimal public reserve price excludes fewer bidder types than under risk neutrality. Moreover, we show that public reserve prices are not optimal as the seller can earn a higher revenue with mechanisms that better leverage the ``attachment effect''. We discuss two such mechanisms: i) an auction with a secrete and random reserve price, and ii) a two-stage mechanism where an auction with a public reserve price is followed by a negotiation if the reserve price is not met. Both of these mechanisms expose more bidders to the attachment effect, thereby increasing bids and ultimately revenue.
  
  %We characterize optimal reserve prices in first-price and second-price auctions with independent private values when bidders are expectations-based loss averse. We find that the optimal public reserve price can be lower than under risk neutrality or risk aversion. Moreover, we show that public reserve prices are not optimal as the seller can earn a higher revenue with a mechanism that better leverages the bidders' ``attachment effect". We discuss two such mechanisms: i) an auction with a secrete and random reserve price, and ii) a two-stage mechanism where an auction with a public reserve price is followed by a negotiation if the reserve price is not met. Both of these mechanisms expose more bidders to the attachment effect, thereby increasing bids and ultimately revenue. 
 	
 	%bidders who are excluded are not exposed to the “attachment effect” and the seller can use other tactics to  secret and random reserve prices raise more revenue than public ones since, by giving every bidder a chance to win, they expose all bidders to the . Furthermore, we draw connections to implications on ratification in other economic contexts.
 	\smallskip
 	
 	\textbf{JEL} \textbf{classification}: D44, D81, D82.
 	
 \end{abstract}
 
 %TCIMACRO{\TeXButton{TeX field}{\thispagestyle{empty}}}%
 %BeginExpansion
 \thispagestyle{empty}%
 %EndExpansion
 \pagebreak
 
 %TCIMACRO{\TeXButton{TeX field}{\setcounter{page}{1}}}%
 %BeginExpansion
 \setcounter{page}{1}%
 %EndExpansion
 
 \section{Introduction\label{intro}}

 Reserve prices are a prevalent tool auctioneers use to raise their expected
 revenue. A reserve price acts as an additional bid placed by the auctioneer since, in
 order to win, a buyer must also outbid the reserve. Thus, a reserve price increases the competitiveness of an auction. %which might lead buyers to bid more aggressively. 
 Yet, this
 comes at a cost for the auctioneer because trade does not happen if no buyer bids at
 least the reserve price. Indeed, a reserve price excludes buyers with
 relatively low valuations from the auction and reduces the overall
 probability of trade. Seminal theoretical contributions by Myerson (1981)
 and Riley and Samuelson (1981) have characterized the revenue-maximizing
 reserve price as the solution to this trade-off between decreasing the
 probability of trade and amplifying competitive pressure. In particular, they show that
 with risk-neutral bidders having independent private values, the optimal
 reserve price coincides with the classical monopoly price; hence, it is (i) deterministic and public, (ii) always higher than the seller's own value, and (iii) under mild conditions on the distribution of
 bidders' values, independent of the number of bidders. %Indeed, the optimal reserve price coincides with the classical monopoly price. 
 
 However, these features are not always empirically observed. For instance, in real-world auctions sellers often use secret reserve prices. A prime example is that of real-estate auctions in the Australian state of Queensland, where prospective buyers are allowed to know whether the seller set a reserve price, but not its exact value.%\footnote{See \url{https://www.qld.gov.au/law/housing-and-neighbours/buying-and-selling-a-property/buying-a-home/ways-to-buy-your-home/buying-at-auction#:~:text=The\%20reserve\%20price\%20is\%20the,will\%20choose\%20to\%20have\%20one.}}
 \footnote{%
 	%While Ashenfelter (1989) provides some early examples of shill bidding in
 	%art auctions, this phenomenon appears to be even more common in online auctions, where sellers can use multiple accounts to bid on their own items; see Grether et al. (2015). 
  Secret reserve prices are also
 	documented by Elyakime et al. (1994) and Li and Perrigne (2003) in timber
 	auctions, and by Bajari and Horta\c{c}su (2004) and Hasker and Sickles (2010) in internet auctions.} 
Moreover, some studies show that empirical reserve prices are often
 significantly lower than what the classical models predict; see Paarsch (1997) and Haile and Tamer (2003). There is also evidence of reserve prices that vary with the number of bidders and auctions with no reserve price at all; see Davis et al. (2011) and Gon\c{c}%
 alves (2013). Overall, this evidence suggests that sellers may face additional
 trade-offs not captured by the classical risk-neutral and/or risk-averse model.\footnote{%
 	With private values, Hu et al. (2010) show that risk aversion can explain
 	low reserve prices in the FPA but not in the SPA; if in addition bidders
 	have interdependent values, Hu et al. (2019) show that risk aversion can
 	explain low reserve prices also in the SPA.} 
 
 In this paper, we analyze reserve prices in first-price auctions (FPA) and second-price
 auctions (SPA) where symmetric bidders have independent private values (IPV) and
 are expectations-based loss averse \'{a} la K\H{o}szegi and Rabin (2006,
 2007, 2009). We derive the revenue-maximizing reserve price for each format
 and highlight how loss aversion modifies the seller's trade-off between
 increasing competitive pressure and reducing the probability of trade. In
 particular, we show that loss aversion can rationalize reserve prices that
 (i) are secret, (ii) vary with the number of bidders, %(iii) differ between
 %auction formats, 
 and (iii) are lower than what the theory predicts for
 risk-neutral and risk-averse bidders.
 
 Section \ref{model} introduces the auction environment and bidders'
 preferences, and describes the solution concept. %We consider a standard
 %symmetric environment with independent private values and analyze two
 %canonical sealed-bid auction formats: the first-price auction (FPA) and the
 %second-price auction (SPA). 
 Following K\H{o}szegi and Rabin (2006), we posit
 that, in addition to classical material utility, a bidder also experiences
 \textquotedblleft gain-loss utility\textquotedblright\ when comparing her
 material outcomes to a reference point equal to her expectations regarding
 those same outcomes, with losses being more painful than equal-size gains are
 pleasant. %For most of the paper, we abstract from loss aversion over money. This simplifying assumption allows us to clearly highlight the novel implications on the optimal reserve price of bidders' motive to reduce (expected) losses.\footnote{As argued by K\H{o}szegi and Rabin (2009), this assumption is reasonable if bidders' income is already subject to large background risk; relatedley, Novemsky and Kahneman (2005) propose that money given up in purchases is not generally subject to loss aversion.}
 
 We apply the solution concept of \textquotedblleft
 unacclimating personal equilibrium\textquotedblright\ (UPE) introduced by K%
 \H{o}szegi and Rabin (2007). According to this concept,
 bidders choose the strategy that maximizes their payoff keeping
 expectations fixed, and the distribution of outcomes so generated must
 coincide with the expectations; hence, when deviating
 from her equilibrium bid, a bidder holds her reference point fixed.\footnote{For other applications of
 	UPE see, for instance, Heidhues and K\H{o}szegi (2008,
 	2014) Karle and Peitz (2014, 2017), Karle and M\"{o}ller (2020), Karle and
 	Schumacher (2017) and Rosato (2016).} As there might be multiple UPEs, %in increasing and symmetric bidding strategies, 
  we assume bidders select their preferred personal equilibrium (PPE) --- the one that maximizes their utility from an ex-ante perspective.

 We begin our analysis in Section~\ref{PPE} by deriving the
 revenue-maximizing public reserve price in the FPA.\footnote{%
 	As shown by Balzer and Rosato (2021), under UPE the FPA and SPA are revenue
 	equivalent; therefore, our results on the reserve price for the FPA carry
 	over to the SPA.} First, we show that the ``attachment effect'' (K\H{o}%
 szegi and Rabin, 2006) is the main driving force behind the bidding behavior of loss-averse buyers. In particular, the higher the probability with which a bidder, in equilibrium, expects to win the auction, the bigger the loss she endures if she ends up losing it. Hence, a bidder has an %stronger 
 incentive to increase her bid, so as to win more often and avoid experiencing the loss. Thus, the attachment effect induces an upward pressure on the equilibrium bidding strategy, ensuring that a bidder's cost from an upward deviation is higher than the benefit. Importantly, the
 attachment effect only affects the incentives of a bidder who expects to win
 the auction with strictly positive probability --- however small --- and is
 thus exposed to potential losses in equilibrium; yet, it does not affect those bidders who abstain from the auction since they do not incur a loss when not winning it.

 The fact that bidders who do not expect to win are not exposed to the
 attachment effect has several implications for the characterization of the
 revenue-maximizing public reserve price. First, it puts downward pressure on
 the optimal reserve price. Indeed, by increasing the reserve price, the
 seller excludes a larger set of bidder types from the auction. As a bidder's
 attachment increases in her type, the higher the marginally excluded type,
 the larger the attachment effect that the seller forgoes. Thus, with
 expectations-based loss aversion, increasing the reserve price is more
 costly for the seller compared to a situation where the attachment effect is
 not present; e.g., with risk-neutral bidders. This finding is especially
 relevant for those empirical studies that first estimate the distribution of
 bidders' values and then use the theoretical insights of Myerson (1981) and
 Riley and Samuelson (1981) to estimate the revenue-maximizing reserve price
 as the minimum bid that excludes all bidders with a ``virtual
 value'' lower than the seller's value. For instance, Paarsch (1997) and Haile and Tamer (2003) find that sellers exclude fewer types than their estimation predicts as revenue maximizing. Yet, as our first result implies, it is theoretically optimal to include in the auction bidders with virtual values lower than the seller's own value because of the attachment effect.  While risk aversion can
 rationalize such low reserve prices in the FPA (see Hu et al., 2010), it cannot
 explain it for the SPA.\footnote{%
 	However, beyond risk aversion, several different explanations for low reserve
 	prices in both the SPA and FPA have been proposed. These
 	include correlated types (Levin and Smith, 1996), interdependent values
 	(Quint, 2017; Hu et al., 2019), endogenous entry (McAfee, 1993; Levin and
 	Smith, 1994; Peters and Severinov, 1997), bidders' selection neglect when
 	sellers are privately informed about the quality of the objects they sell (Jehiel and Lamy, 2015), level-k bidders (Crawford et al., 2009) and
 	taste projection (Gagnon-Bartsch et al., 2021).} 
 
Furthermore, the optimal public reserve price varies with the number of bidders. Indeed, the more bidders are present, the less optimistic each of them is about her chances of winning, which reduces their attachment. Yet, this does not imply that the reserve price always increases in the number of bidders. Indeed, when raising the reserve price, the seller forgoes the attachment effect of the marginal type. However, with already many bidders participating, adding an extra one reduces this cost, leading to more exclusion.\footnote{Menicucci (2021) obtains a similar result in the classical IPV risk-neutral model when the bidders' virtual
 	values are not monotone; in contrast, our result holds also for the regular case of increasing  virtual
 	values.} With risk aversion, instead, the optimal reserve price (in the FPA) is decreasing in the number of bidders; see Vasserman and Watt (2021). 
 
 %Third, the fact that the attachment effect does not operate on excluded
 %bidders, who rationally expect to lose the auction with %certainty, gives rise to the question: \textit{Can the seller employ tactics to increase the attachment of those types?} As we show in Section~\ref{RPPE}, the answer to this question is affirmative. In particular, we characterize two such tactics that deliver a strictly larger revenue than an auction with a public reserve price.
 The fact that the attachment effect does not operate on those bidders excluded by a public reserve price suggests that a seller could raise an even higher revenue by exposing more bidders to this effect. In Section \ref{RPPE} we show that this intuition is correct. In particular, we characterize two tactics whereby a seller can expose almost all bidders to the attachment effect, resulting in a strictly larger revenue than an auction with a public reserve price.

 In Subsection~\ref{sec:Secret and Random} we show that
 secret and random reserve prices are revenue superior to public and
 deterministic ones. To see why, notice that with a secret reserve price each bidder type expects to win the auction with
 strictly positive --- albeit potentially arbitrarily small --- probability.
 In such an auction, therefore, every bidder is exposed to potential losses and thus has
 an incentive to bid more aggressively in order to avoid them. Hence, by
 transforming the public reserve price into a secret one, the seller can
 ensure that every bidder experiences the attachment effect, which enhances
 revenue. By doing so, however, the seller also reduces the competitive
 pressure on the buyers' side, which could potentially harm revenue since
 those low-type bidders excluded under a public reserve would be
 participating now. Yet, the seller can choose a distribution for the
 (secret) reserve price that puts large probability mass on relatively high
 prices and arbitrarily small mass on low ones. Such a distribution ensures
 that, while the seller exposes every bidder to the attachment effect, the
 competitive pressure is almost the same as under a public reserve price. 
 
 %As argued by Bajari and Horta\c{c}su (2004), secret reserve prices are
 %common in internet auctions.\footnote{%
 %	Secret reserve prices on Ebay are deterministic; yet, from the bidders'
 %	perspective, the secrete reserve price might appear random if they do not observe
 %	how precisely the seller chooses it. Indeed, the distribution that we characterize for the optimal
 %	secret reserve price puts a large mass on the upper bound of the
 %	support and arbitrarily little mass everywhere else, consistent with the notion of
 %	bidders expecting the seller to make small ``mistakes''. In timber auctions, secret
 %	reserve prices are ``truly'' random, as argued by Li and
 %	Perrigne (2003).} %
  Thus, expectations-based loss aversion provides a novel rationale for
 secret \emph{and} random reserve prices.\footnote{An indirect way of implementing a secret and random reserve price is via ``shill bidding'', a prominent albeit often illegal practice in real-world auctions whereby a dummy buyer submits pre-specified bids on behalf of the seller; see Ashenfelter (1989).} This result is reminiscent of those in Heidhues and K\H{o}szegi (2014) and Hancart (2022), who characterize the optimal pricing strategy for a monopolist selling to an expectations-based loss-averse buyer. In line with the findings of Azevedo and Gottlieb (2012), who showed that risk-neutral sellers benefit from offering gambles to consumers exhibiting prospect-theory preferences, these papers find that the monopolist benefits from using random prices. In particular, Heidhues and K\H{o}szegi (2014) show that if the seller has sufficient commitment power, a stochastic pricing scheme featuring low, variable sale prices and a high, sticky regular price yields more revenue than posting a single price. Our secret and random reserve price scheme has similar features, but differs from their characterization since we consider an environment with multiple, privately-informed buyers. Nonetheless, we are able to draw a connection between the optimal reserve price and the optimal monopoly pricing scheme with expectations-based loss-averse buyers that is analogous to the well-known one for risk-neutral buyers.

 In Subsection~\ref{sec:price positing} we show that the seller can achieve a higher revenue than what is achievable with a public reserve price by employing a simple two-stage mechanism. In this mechanism, the seller first runs an auction with a public reserve price; then, if the reserve price is not met, with some probability, the seller posts a price that would be accepted by those bidders whose types are below the marginally excluded one. In this way, the seller exposes also the bidder type marginally excluded from the initial auction to the attachment effect; this, in turn, pushes the marginal type to bid more aggressively, thereby increasing the overall revenue. %This mechanism not only generates a higher revenue compared to committing to a public reserve price, but it generates the same revenue as secrete reserve prices in the limiting case where the interval of types that accept the posted price converges to the threshold type. However, for a sufficiently high type it revenue dominates public reserve prices. 

 %For the same reason, and differently from UPE, the optimal reserve price always excludes more types compared to the risk-neutral benchmark. This is %the case because, under CPE, even the marginally excluded bidder is exposed to potential losses (in the good's dimension) when participating in the %auction. To induce a given threshold type to participate, therefore, the seller has to set a rather small reserve price. In turn, the seller's %forgone revenue from increasing the threshold type are smaller than under risk neutrality and she optimally excludes more types.  

 %Second, the revenue-maximizing
 %reserve price is larger than that under UPE.

 Section~\ref{conclusion} concludes the paper by summarizing the results of
 our model and discussing some further implications. All proofs are relegated
 to Appendix~\ref{proofs}.

\section{The Model} \label{model}

{\small 
%\textbf{basic outcomes can be defined both in the SPA and the FPA; SPA, just say the outcomes are in this interval} 
}

In this section, we introduce the auction environment and bidders'
preferences, and provide a formal definition of our solution concept (UPE) in the context of sealed-bid auctions.

\subsection{Environment}

A seller auctions off an item to $N\geq 2$ bidders via a sealed-bid auction. Each bidder $i\in \left\{
1,2,...,N\right\} $ has a private value $t_{i}$ independently drawn from the support $\left[ \underline{t},\overline{t}\right] 
$, with $\overline{t}>\underline{t}= 0$, according to
the same cumulative distribution function $F$.\footnote{We normalize $\underline{t}= 0$ to simplify the exposition. Moreover, under this assumption, a seller facing risk-neutral bidders would always choose a non-trivial reserve price; i.e., there are no corner solutions.} We assume that $F$ is continuously
differentiable, with strictly positive density $f$ on its support. Further,
we impose the standard assumption that $F$ has a monotone hazard rate; i.e,  $ 
\frac{f(x)}{1-F(x)}$ is increasing for all $x \in \left[ \underline{t},\overline{t}\right]$. This, in turn, implies that bidders' ``virtual values'' are increasing; i.e., $V(t_i)\equiv t_i-\frac{1-F(t_{i})}{f(t_{i})}$ is increasing in $t_i$. The seller has a commonly-known value $t^S \in [ 0, \bar t)$. 

We
consider two canonical selling mechanisms: the
first-price sealed-bid auction (FPA) and the second-price sealed-bid auction (SPA). We restrict
attention to symmetric equilibria in increasing strategies; in such
equilibria, the bidder with the highest type wins the auction, conditional on placing a bid above the reserve price.\footnote{Throughout the paper, we restrict attention to symmetric (i.e., non discriminatory) auction mechanisms; for a recent analysis of asymmetric auctions with expectations-based loss-averse bidders, see Muramoto and Sogo (2022).} Let $F_{1}$ denote the cumulative distribution function of the
highest order statistic among $N-1$ draws, and denote by $f_{1}$ its corresponding density. Finally, let $r_{RN}$ denote the revenue-maximizing reserve price with risk-neutral bidders, and notice that $r_{RN}>0$ since $\underline{t}=0$.

%\begin{definition}
%Consider a selling mechanism and suppose that $q(t )=0$ for bidder-type $t $. Then, we say that bidder type-$t $ agent is excluded from the mechanism.
%\end{definition} 

\subsection{Bidders' Preferences and Solution Concept}

Consider a bidder participating in either an FPA or an SPA; depending on her bid and her opponents' ones, she might either win the auction ($q=1$) in which case she receives the item and pays a price $p \in \mathbb{R}_+$, or lose the auction ($q=0$) in which case she does not obtain the good and pays nothing. We assume bidders have expectations-based reference-dependent preferences as formulated by K\H{o}szegi
and Rabin (2006, 2007, 2009). Accordingly, the utility of bidder $i$ with type $t_i$ has two components. First, her material utility is given by $q(t_i-p)$, with $q\in\{0,1\}$. Second, the bidder also derives psychological utility from
comparing the realized outcome to a reference outcome given by
her recent expectations (probabilistic beliefs).\footnote{%
Banerji and Gupta (2014), Rosato and Tymula (2019) and Eisenhuth and Grunewald
(2020) provide experimental support for the K\H{o}szegi and Rabin's
model in the context of sealed-bid auctions.} Hence, given an outcome $\left( q,p\right) $ and a deterministic reference point $\tilde{r} \in \{0,1\}
%\times \mathbb{R} 
$, a bidder's total utility is{\small 
\begin{equation}
U\left[ \left( q,p\right) | \tilde{r},t_i \right]
=q(t_{i}-p)+\mu \left( qt_{i}-\tilde{r}t_{i}\right) 
\label{eq14}
\end{equation}%
}where{\small 
\begin{equation*}
\mu \left( x\right) =\left\{ 
\begin{array}{ccc}
\eta   x & \text{if} & x\geq 0 \\ 
\eta  \lambda   x & \text{if} & x<0%
\end{array}%
\right.
\end{equation*}%
}is \textit{gain-loss utility}, with\ $\eta \geq 0$ and $\lambda  >1$. The
parameter $\eta  $ captures the weight a bidder attaches to
gain-loss utility while $\lambda  $ is the coefficient of loss aversion.\footnote{To clearly highlight the implications of the attachment effect on bidding incentives, we depart from the original formulation of K\H{o}szegi
and Rabin (2006, 2007, 2009) by considering buyers who are loss averse only
with respect to their value for the item, but not with respect to the price they might pay; in other
words, we assume buyers are risk neutral over money. As argued in K\H{o}szegi
and Rabin (2009), this assumption
is reasonable if buyers' income is already subject to large background risk; relatedly, Novemsky and
Kahneman (2005) propose that money given up in purchases is not generally subject to loss aversion.}
%Moreover, according to (%
%\ref{eq14}), a bidder assesses gains and losses separately over each
%dimension of consumption utility, with different gain-loss parameters.\footnote{%
	%We allow for different parameters of gain-loss utility on
	%the good and money dimensions because the two have different implications
	%for bidding in auctions. In particular, our formulation is rich enough to
	%capture situations where bidders are loss averse only regarding the
	%consumption dimension. Such case applies if bidders' income is subject to
	%large background risk, as argued by K\H{o}szegi and Rabin (2009); in a
	%similar vein, Novemsky and Kahneman (2005) argue that money given up in
	%purchases may not be subject to loss aversion.}
% this is consistent with much of the
%experimental evidence commonly interpreted in terms of loss aversion.%
%\footnote{%
%In particular, this feature can rationalize the endowment effect observed in
%many laboratory experiments (see Kahneman et al\textit{.}, 1990; 1991);
%indeed, the common explanation of the endowment effect is that owners
%perceive giving up an object as a painful loss that counts more than the
%money received in exchange.}

Because in many situations expectations are stochastic, K\H{o}szegi and
Rabin (2006, 2007, 2009) allow for the reference point to be described by a
distribution $H$ over the possible values of $\tilde{r}$; then, fixing $H$, a bidder's total utility from the outcome $\left(
q,p\right) $ can be written as%
\begin{equation*}
U\left[ \left( q,p\right) | H  ,t_i \right]
=q(t_{i}-p)+\int_{\tilde{r}}\mu \left( qt_{i}-\tilde{r}t_{i}\right) dH{\left( \tilde{r}\right)}
  .
\end{equation*}

In words, a bidder compares the realized outcome
to all possible outcomes in the reference lottery, each one weighted by its respective
probability.
 
A bidder learns her type before submitting a bid and, hence, maximizes her
interim expected utility. If the distribution of the reference point is $%
H $ and the distribution of consumption outcomes
is $G=\left( G^{g},G^{m}\right) $, the interim expected utility of a bidder
with type $t_{i}$ is%
\begin{equation*}
EU\left[ G|H,t_i\right] =\int_{\left\{ q,p\right\}
}\int_{\left\{ \tilde{r} \right\} }U\left[ \left( q,p\right) | 
\tilde{r}   , t_i  \right] dH\left( \tilde{r} \right) dG%
\left( q,p\right) .
\end{equation*}

A strategy for bidder $i$ is a function $\beta _{i}:\left[ \underline{t},%
\overline{t}\right] \rightarrow \mathbb{R}_{+}$. Fixing all other bidders'
strategies, $\boldsymbol{\beta}_{-i}$, the bid of bidder $i$ with type $%
t_{i} $, $\beta _{i}(t_{i}),$ induces a distribution over the set of final
consumption outcomes. Let $\Gamma \left( \beta _{i}(t_{i}),\boldsymbol{\beta 
}_{-i}\right) $ denote this distribution. %In a sealed-bid auction,
%uncertainty is resolved after all bids are submitted. Thus, holding her
%opponents' strategies fixed, a bidder's strategy affects the distribution
%over consumption outcomes. As pointed out by %K\H{o}szegi and Rabin (2007),
%when a person makes a committed decision long before outcomes occur, she
%affects the reference point with her choice so that $G\equiv H$.  
 According to K\H{o}szegi and Rabin (2007), when a
decision is made shortly before outcomes realize, the reference point is
fixed by past expectations; then, when the decision maker chooses the bid that maximizes her
expected utility, she takes the reference point as given. Being fully
rational, therefore, she can plan to submit a bid only if she is willing to
follow it through, given the reference point determined by the expectation
to do so. This is what K\H{o}szegi and Rabin (2007) call unacclimating
personal equilibrium (UPE):

\begin{definition}
\label{UPE} A strategy profile $\boldsymbol{\beta}^{\ast }$ constitutes an
Unacclimating Personal Equilibrium (UPE) if for all $i$ and for all $t_{i}$:%
\begin{equation*}
EU\left[ \Gamma \left( \beta _{i}^{\ast }(t_{i})\boldsymbol{,\beta }%
_{-i}^{\ast }\right) |\Gamma \left( \beta _{i}^{\ast }(t_{i})\boldsymbol{%
,\beta }_{-i}^{\ast }\right) ,t_{i}\right] \geq EU\left[ \Gamma \left( b%
\boldsymbol{,\beta }_{-i}^{\ast }\right) |\Gamma \left( \beta _{i}^{\ast
}(t_{i})\boldsymbol{,\beta }_{-i}^{\ast }\right) ,t_{i}\right]
\end{equation*}%
for any $b\in \mathbb{R}_{+}$.
\end{definition}

Thus, if a bidder deviates to a
different bid, her reference point does not change. Notice that there
might exist multiple UPEs; that is, multiple bids that the bidder is
willing to follow through. In this case, following K\H{o}szegi and Rabin
(2006, 2007), we assume that the bidder selects the UPE that provides her
with the highest expected utility. Hence, bidders
play according to their (symmetric) Preferred Personal\ Equilibrium (PPE).\footnote{%
See also Heidhues and K\H{o}szegi (2014), Rosato (2016), Freeman (2019), and
Balzer and Rosato (2021).}

\section{Deterministic and Public Reserve Price\label{PPE}}

This section characterizes the optimal public reserve price. Let $q(t)$ denote the probability with which, in equilibrium, a type-$t$ bidder wins the auction. In the FPA, without a reserve price, the highest bidder wins
the good and pays her bid. Hence, in a symmetric equilibrium, it
holds that $q(t)=F_{1}(t)$ for all $t\in \lbrack \underline{t},\bar{t}]$. However, with
a (binding) reserve price $r$, there is a threshold type $t_{r}$ such that $%
q(t)=0$ for all $t\in \lbrack \underline{t},t_{r})$ and $q(t)=F_{1}(t)$ for $%
t\in \lbrack t_{r},\bar{t}]$; that is, all bidders with types below $t_{r}$ prefer not to
participate in the auction.

Fix a symmetric and increasing bidding strategy, $\beta _{I}:[\underline{t},%
\bar{t}]\mapsto \mathbb{R}_{+}$. Moreover, fix $r$ and the implied $t_{r}$,
to be determined shortly, and consider a type-$t$ bidder who mimics a \emph{%
larger} type $\tilde{t}>t$.\footnote{As shown by Balzer and Rosato (2021), in the PPE bidders' upward incentive constraints are the binding ones.} With a slight abuse of notation, denote her expected payoff by $EU(\tilde{t},t)$; this is given by
\begin{eqnarray}
EU(\tilde{t},t) &=&q(\tilde{t})(t-\beta _{I}(\tilde{t}))-\eta \lambda
(1-q(\tilde{t}))q(t)t+\eta (1-q(t))q(\tilde{t})t ,   \label{EUOPFPA}
\end{eqnarray}%
where $q(x)=F_{1}(x)$ if $x\geq t_{r}$ and $q(x)=0$ otherwise.

The first term on the right-hand side of (\ref{EUOPFPA}), $q(\tilde{t})(t-\beta _{I}(\tilde{t}))$, is a bidder's
expected material payoff. The
second and third terms represent the (expected) gains and losses for a bidder who planned to bid $\beta_I(t)$, hence expecting to
win with probability $q(t)$, but then deviates and bids $\beta_{I}(\tilde{t})$, and thus wins with probability $q(\tilde{t})$. Whenever she loses, the bidder experiences a loss of $\eta \lambda
q(t)t$ weighted by the probability of losing the auction, $1-q(\tilde{t}%
)$. Similarly, given that the bidder expected to lose the auction with
probability $1-q(t)$, if she ends up winning it she experiences a gain of $%
\eta (1-q(t))t$ weighted by the probability with which that gain occurs, 
$q(\tilde{t})$.  

Now consider a bidder with type $t\in \lbrack \underline{t},t_{r})$. In
equilibrium, such a bidder does not want to mimic the threshold type $t_{r}$
implying that 
\begin{eqnarray}
EU(t_{r},t) &\leq &EU(t,t) \Leftrightarrow \notag \\
F_{1}(t_{r})(1+\eta )t &\leq &F_{1}(t_{r}) r.  \label{PPEmarginaltype}
\end{eqnarray}

To understand condition (\ref{PPEmarginaltype}) note that, in equilibrium,
bidders whose types are in $[\underline{t},t_{r})$ do not participate in the auction and thus expect
to win the good with zero probability. If one of these bidders deviates
and mimics type $t_{r}$,  she then wins the auction with probability $F_{1}(t_{r})$. Thus, her expected gains from deviating entail a material gain of $F_{1}(t_{r})t$ and
a psychological gain of $F_{1}(t_{r})\eta t$, since she
expected to lose the auction for sure; hence, the terms on the left-hand side of (\ref%
{PPEmarginaltype}) represent the benefits from deviating and submitting a bid equal to the reserve price. The right-hand side of (\ref%
{PPEmarginaltype}) captures the expected costs from such a deviation --- the increase in the expected payment, $F_1(t_r)r$.  Letting $t\rightarrow t_{r}$ from below, and making (\ref{PPEmarginaltype})
hold with equality, we obtain the following relationship
%\footnote{%In the PPE, bidders' upward incentive constraints bind.} 
between the reserve
price, $r$, and the type of the marginal bidder: 
\begin{equation}
r= \left( 1+\eta  \right)t_{r}.\footnote{Note that $r>t_r$ follows from our assumption of no loss aversion over money. If a bidder is also loss averse over money and expects to not pay anything, she experiences a loss of $\eta \lambda r$ when winning at the reserve price; in this case the mapping between the reserve price and the marginal type becomes $r= \left(1+\eta\right)t_{r}/\left(1+\eta \lambda \right)$.}\label{reserve}
\end{equation}

For a given type $t^{\prime }$, in the following we denote the solution to (\ref%
{reserve}) by $r(t^{\prime })$ (i.e., the reserve price assuring that $%
t_{r}=t^{\prime }$). Using the relationship between the threshold type $%
t_{r} $ and the reserve price, we then apply the standard logic from auction theory:
equilibrium behavior shapes the bidding function up to a constant, which is pinned down by type $t_{r}$'s expected payment, $F_{1}(t_{r})r$. The next lemma
formally states a bidder's expected payment in the PPE for a given $t_{r}$. 

\begin{lemma}
	\label{L1} Consider the PPE of an FPA with deterministic
	reserve price. Let $t_{r}$ be the
	lowest type that receives the good with strictly positive probability. The expected payment from a bidder with type $t \geq t_{r}$ is: 
	%\begin{equation}

	\begin{align} \label{Ep}
		F_{1}(t)\beta _{I}^{\ast }(t)  &= \int_{t_{r}}^{t} %
			[1+\eta \lambda F_{1}(x)+\eta (1-F_{1}(x))] %
		f_{1}(x)xdx  +  (1+\eta )F_{1}(t_{r})t_{r} 
	\end{align}
	%\end{equation}

	and 0 for any $t< t_{r}$.
\end{lemma}

Notice that expressions (\ref{reserve}) and (\ref{Ep}), and hence $\beta _{I}^{\ast }(t)$, reduce to their well-known risk-neutral analogues if $\eta  =0$. For $\eta>0$ and $t\geq t_r$, instead, $\beta_I^*(t)$ is strictly larger than its risk-neutral counterpart since buyers have an additional incentive to raise their bids in order to try to win more often and reduce their expected losses.\footnote{As shown in Balzer and Rosato (2021), when buyers are also loss averse over money, they might bid less than in the risk-neutral benchmark if their type is sufficiently low.} This is the attachment effect: the larger the probability with which a bidder expects to win, the bigger the loss she feels if she loses, and hence the stronger her incentive to raise her bid. Hence, because of the attachment effect, bidders' willingness to pay endogenously depends on their expectations. In what follows, %we elaborate on this point by relating bidders' marginal willingness to pay to their equilibrium behavior (bid functions).
we will focus on the relationship between the attachment effect and the bidders' marginal willingness to pay, and its implications for the seller's choice of the optimal reserve price. 

Consider a type-$t$ bidder who expects to win the auction with probability $q$. How much does this bidder value an increase, $\Delta q>0$, in her probability of winning?
If her probability of winning increases by $\Delta q$, the bidder obtains the good more often; hence, she makes a material gain equal to $t\Delta q$. Moreover, the bidder's chances of enjoying a psychological gain increase, which she values at $\eta (1-q)\Delta q t$, given that she expected to lose with probability $1-q$. Similarly, by winning more often, the bidder's chances of experiencing a loss are also reduced; she values this reduction in the probability of making a loss at $\eta \lambda q \Delta q t$. Adding up all these terms, the bidder's marginal willingness to pay for such an increase in the probability of obtaining the good is equal to

\begin{equation}\label{MWTP}
    \underbrace {t\Big(1+\eta \lambda q+\eta(1-q) \Big)}_{MWTP(t;q)}\Delta q.
\end{equation}

%$MWTP(t;q )=	t\Big(1+\eta \lambda q+\eta(1-q) \Big)$, and depends on the her expected winning probability. To highlight how expectations shape the MWTP, we distinguish the two following cases:  
%\begin{eqnarray}\label{MWTP}
 %\Delta q \times MWTP( t; q,)= \begin{cases} 
	%	t\Big(1+\eta \lambda q+\eta(1-q) \Big)	\Delta q     ,& \quad \textit{if } q  > 0 \\
	%	t (1+\eta )\Delta q    ,& \quad \textit{if } q =0
	%\end{cases}
%\end{eqnarray}

It easy to see that $MWTP(t;q )$ increases not only in the bidder's type $t$, but also in the probability $q$ with which she already expects to win since $\lambda>1$; in particular, a bidder who expects to never get the good (i.e., $q=0$) is willing to pay less for a given $\Delta q$ than a bidder who expects to obtain the good with strictly positive probability (i.e., $q>0$). This is because only a bidder who rationally expects to win the good experiences a loss when not winning. %Thus, such a bidder values an increase in the winning probability more than a bidder who expects to lose for sure and thus to never suffer a loss.

%Comparing the first and second line of \eqref{MWTP}, we see that the term  $\eta \lambda q \Delta q t$ only appears in the first one. This term is the reduction in the probability of making a loss (when winning with $\Delta q$ more often) times the loss bidder-$t$ who expected to win with probability $q$ experiences when losing. The third term in the first line, $\eta (1-q)\Delta q t$,  is the increase in the probability of winning times the gain a type-$t$ bidder, who expected to lose with probability $1-q$, experiences when winning. The respective term, with $q=0$, also shows up in the second line of \eqref{MWTP}. Indeed, a bidder who expected to lose the auction with certainty, $q=0$, experiences a gain when winning. 

Consider now a bidder with a type arbitrarily below the threshold type, $t_r$, i.e., $t \rightarrow t_r$. In equilibrium, as under risk neutrality, the seller charges her an expected payment equal to her entire willingness to pay. That is, 
\begin{equation*}
 F_1(t_r) r= \int_0^{t_r} MWTP(t_r;0) f_1(s) ds = F_1(t_r)(1+\eta)t_r.
\end{equation*} 

In other words, in equilibrium the reserve price makes the threshold type indifferent between participating or not. 
  
Next, consider a type-$t$ bidder with 
$t >t_r$. In equilibrium, $q(t)=F_1(t)$ and thus $\Delta q=f_1(t)$; hence, such a bidder's $MWTP(t;q)$ from mimicking a slightly larger type (and thus increasing her winning probability by $f_1(t)$) must equal the increase in the expected payment, $(F_1(t)\beta_I^*(t))'$, from such a deviation.
That is, $F_1(t)\beta_{I}^*(t)=\int_{t_r}^t MWTP(s;F_1(s))f_1(s) ds+F_1(t_r)r$, i.e., \eqref{Ep}. 

The preceding discussion shows that the attachment effect, which is revenue enhancing for the seller, has a bigger influence on  the behavior of those bidder types who expect to win with strictly positive probability (i.e., $t>t_r$) than on the behavior of the threshold type who bids the reserve price. This occurs because the seller can charge the threshold type a price capturing only to the additional gain from winning, but not the benefit from avoiding losses. This asymmetry is crucial for the determination of the optimal reserve price under loss aversion $r^*$  and, in turn, the degree of bidder exclusion, as the next result shows.
%The next result compares the effect of the optimal reserve price under loss aversion, $r^*$, on the probability of a trade with that under risk neutrality.

\begin{proposition}\label{P1r}
	\label{P1} The optimal threshold type in the
	FPA, $t_r^*$, is smaller than that under risk
	neutrality, $t^{RN}$. Moreover, for sufficiently large $\lambda$, $t_r^{\ast}=r^{\ast}=0$, even if $t^S>0$.
\end{proposition}

Hence, with expectations-based loss-averse bidders, the seller optimally excludes fewer types than in the risk-neutral benchmark. To see the intuition, consider the seller's trade-off when setting the optimal threshold type; she chooses $t_r$ in order to maximize her expected profit given by
\begin{equation}
N \int_{t_r}^{\bar{t}} F_1(s)\beta_I^{\ast}(s) f(s)ds + F(t_r)^N t^S,
\end{equation}
where the integral term represents the expectation over buyer's expected payments, as given in equation~(\ref{Ep}), and the last term is the seller's payoff if no trade takes place.
%The seller's objective function is the expected revenue plus the opportunity cost from selling the good, that is, $F(t_r)^N t^S$.
The derivative of the seller's profit with respect to $t_r$ takes the following form:
 
\begin{equation}\label{myerson}
 N  \Big[	(1+\eta) f(t_r)	\left(    \frac{1-F(t_r)}{f(t_r)}-t_r  \right)+f(t_r) t^S -   \eta(\lambda-1)(1-F(t_r))     f_1(t_r)t_r  \Big]F_1(t_r).
\end{equation}
Note first that, because this is a symmetric environment, all terms in the first-order condition are multiplied by the number of bidders, $N$.
The first term is proportional to the (negative of the) virtual value of the threshold type, and it captures the standard trade-off between raising competitive pressure and risking not to sell the good at all. The second term captures the standard effect of the seller's opportunity cost of selling: reducing the probability of trade by marginally raising the threshold type is less costly the higher is the seller's own value for the good. Finally, the last term captures a novel trade-off due to expectations-based loss aversion. Indeed, using \eqref{MWTP}, it is easy to see that this term equals $( MWTP(t_r,0 ) - MWTP(t_r,F_1(t_r) ))f_1(t_r)$; i.e., the difference between the threshold type's marginal willingness to pay when expecting to lose for sure, $MWTP(t_r,0 )$, and her marginal willingness to pay when expecting to win with probability $F_1(t_r)$, $MWTP(t_r,F_1(t_r) )$, multiplied by the change of the winning probability when marginally increasing the threshold type, $f_1(t_r)$ . If the seller raises the threshold type, she transforms an interior type (a type that expects to win with probability $F_1(t_r)$) into the new threshold type. By doing so, she loses the attachment effect of that new threshold type. %, $MWTP(t_r,F_1(t_r) )$. % and only obtains the effect on the increased absolute willingness to pay of the new threshold type, $MWTP(t_r,0)   $.  
 
Therefore, the attachment effect creates an additional cost for the seller of raising the threshold type and thus excluding more types. Moreover, this additional cost increases in the degree of bidders' loss aversion. Indeed, for a large enough $\lambda$, the optimal reserve price is equal to zero even though the seller's value for the item is strictly positive. %Furthermore, the larger the threshold type, the larger this additional cost is since the attachment effect is stronger for higher bidder types who expect to win more often. 
As a consequence, the optimal threshold type with loss aversion is lower than under risk neutrality, which in turn implies less bidder exclusion.

By relating the optimal threshold type under expectations-based loss aversion to its risk-neutral counterpart, Proposition~\ref{P1r} is relevant not only from a theoretical point of view but also from an applied one. Indeed, several empirical papers (e.g., Paarsch, 1997; Haile and Tamer, 2003) use bids submitted in actual auctions to estimate the distribution of bidders' values and, given these estimates, conclude that sellers in the field set reserve prices, and hence threshold types, that are lower than Myerson (1981)'s optimal one (i.e., the one that equates a bidder's ``virtual value" with the seller's value), resulting in too little exclusion. Proposition~\ref{P1r}, however, shows that such seller behavior is consistent with profit maximization if bidders are expectations-based loss averse.

 The next proposition describes how the optimal threshold type, and hence the reserve price, vary with the number of bidders.

\begin{proposition}
\label{N}  The probability of no trade and the
reserve price depend on the number of bidders. For any $t^S\geq 0$, the no-trade probability and the optimal reserve price %weakly 
increase in $N$ if and only if $ln(F(t_r^*))<-1/(N-1)$. Moreover, if $t^S=0$ (resp. $t^S>0$), as $N\rightarrow \infty $, the no-trade probability converges to (resp. is strictly lower than) its risk-neutral counterpart.
%Moreover, for any $t^S \geq 0$, if $N \rightarrow \infty $ the optimal reserve price is larger than under risk neutrality.  
\end{proposition}

From Proposition \ref{P1r}, we already know that $t^*_r\leq t^{RN}$. Then, to see the intuition behind Proposition \ref{N}, suppose first that $t^S=0$ and recall that the optimal threshold type in the risk-neutral benchmark is independent of $N$. With expectations-based loss aversion the optimal level of bidder exclusion depends also on the attachment to which the marginal type is exposed, as captured by the last term in expression (\ref{myerson}). It is easy to see that this term depends on $f_1(t_r)$ and hence on $N$. In particular, when $f_1(t_r^*)$ decreases in $N$, which happens if and only if $ln(F(t_r^*))<-1/(N-1)$, the seller's cost of raising the threshold type due to the forgone attachment effect decreases in the number of bidders; consequently, she excludes more types.

%Recall that $t_r^*$, which depends on $N$, is lower than $t^{RN}$, which is itself independent of $N$. Thus, the left-hand-side of the condition $ln(F(t_r^*))<-1/(N-1)$ is strictly negative for any $N$. In contrast, the right-hand-side of the condition is increasing in $N$ and converges to zero as $N$ grows large. In turn, if $N$ is sufficiently large, it is the case that $ln(F(t_r^*))<-1/(N-1)$ and the threshold type increases in $N$.  
%To see the intuition for this result suppose first that $t^S=0$, and note that $f_1(t_r^*)$ decreases in $N$ if and only if $ln(F(t_r^*))<-1/(N-1)$. Under this condition, the seller's cost from raising the threshold type due to the forgone attachment effect decreases in the number of bidders. Consequently, she excludes more types. 
In line with this intuition, as $N\rightarrow \infty $ every bidder type except $\bar{t}$ expects to win the
auction with (almost) zero probability (i.e., if $N$ is sufficiently large $f_1(t_r^*)$ decreases in $N$ and converges to zero); in turn, the attachment effect of any bidder type but the highest one becomes negligible. In this case, the seller does not affect any
such type's expectations of winning when raising the reserve price, and thus sets the same threshold type as in the risk-neutral benchmark. However, the limit probability of no trade under loss aversion is lower than its risk-neutral counterpart if $t^S>0$. The reason is that loss-averse buyers bid more aggressively than risk-neutral ones; i.e.,  the seller raises more revenue from a loss-averse buyer than from a risk-neutral buyer with the same type. Hence, the seller has a weaker incentive to exclude them.

Finally, note that with risk-averse bidders the optimal reserve price depends on the number of bidders in the FPA, but not in the SPA; see Hu et al. (2010) and Hu (2011). In contrast, all the results for loss-averse bidders derived in this section hold also for the SPA. Indeed, Balzer and Rosato (2021) show that both auction formats yield the
same expected revenue to the seller. Moreover, since the threshold type $%
t_{r}$ does not face any risk in her payment conditional on winning, the
relationship between $r$ and $t_{r}$ satisfies (\ref{reserve}) also in the SPA. Hence,
the optimal reserve price and the optimal threshold type in the SPA coincide with those in the FPA.

\section{Exposing More Bidders to the Attachment Effect}\label{RPPE}

As the previous section highlighted, the seller benefits from the attachment effect, as it pushes buyers to bid more aggressively; however, the excluded buyers are not exposed to this effect. %This suggests that the seller has an incentive to expose as many buyers as possible to the attachment effect. 
In this section, we investigate tactics that expose more bidders to the attachment effect, thereby boosting the seller's revenue. We focus on two such tactics. The first one is a standard auction with a secrete and random reserve-price regime. Heidhues and K\H{o}szegi (2014) and Hancart (2022) show that for a monopolist facing a loss-averse buyer, committing to a stochastic pricing strategy yields a higher revenue than posting a single price, as in this way the seller ensures that the consumer is fully attached to the good. In our setting with multiple buyers, a public (and deterministic) reserve price corresponds to a single posted price whereas a secret and random reserve price corresponds to a stochastic pricing strategy. The second tactic is an auction with a public reserve price followed by a take-it-or-leave-it (TIOLI) negotiation if the reserve price is not met.%maybe we should call it pooling as we do not want to get into dynamics with reference point updating

\subsection{Random and Secret Reserve Prices}\label{sec:Secret and Random}

 In what follows, we first assume that every bidder type is exposed to the attachment effect --- even those excluded --- and solve for the revenue-maximizing auction under this assumption. This exercise leads to a strict upper bound on the seller's revenue. We then relax the assumption that all bidders are exposed to the attachment effect, and show that an auction mechanism using secret and random reserve prices achieves a revenue arbitrarily close to this upper bound.

%In what follows, we construct a strict upper bound on the seller's revenue and maximize that bound. After that, we focus on the implementation and describe two tactics that allow the seller to achieve a revenue arbitrarily close to that upper bound in the original model.\bigskip

 Start by considering a pseudo auction where every bidder type is exposed to the attachment effect that arises in a symmetric auction without exclusion; that is, let $\widehat{MWTP}(t;F_1(t))\equiv MWTP(t;q)|_{q=F_1(t)} $ (as defined in \eqref{MWTP}), even for bidder types that are excluded, i.e., bidders for whom $q(t)=0$. We then calculate the %equilibrium
 bidding function, $\hat{\beta}$, for this hypothetical situation.
Given this pseudo bidding function, the seller maximizes her revenue by using a public reserve price $\hat r$ that, if $t^S=0$, excludes the same set of types as the optimal auction under risk neutrality; this is because if buyers were to bid according to $\hat \beta$, %instead of their original bidding functions, 
the seller would not face the additional cost from the foregone attachment effect of the marginally excluded type.

 Consider now a bidder with type $t=t_{\hat r}$ who in equilibrium bids $\hat r$; i.e., the threshold type. As under risk neutrality, in equilibrium her expected payment equals her entire willingness to pay: 
\begin{equation*}
 F_1(t_{\hat r}) \hat r= \int_0^{t_{\hat r}} \widehat{MWTP}(t_{\hat r};F_1(s)) f_1(s)ds = [1+\eta+\eta(\lambda-1) F_1(t_{\hat r})/2]F_1(t_{\hat r})t_{\hat r}.
\end{equation*} 
\smallskip
In other words, the reserve price makes the threshold type indifferent between participating or not.

Next, consider a type-$t$ bidder with
$t >t_{\hat r}$. In this hypothetical equilibrium, $q(t)=F_1(t)$ and thus $\Delta q=f_1(t)$; hence, such a bidder's (pseudo) marginal willingness to pay to increase her winning probability by $f_1(t)$ must equal the increase in the expected payment, $(F_1(t)\hat \beta(t))'$, from such deviation. Thus,  
\begin{eqnarray}\label{upper_bound_bidding}
F_1(t)\hat \beta(t)=\int_{t_{\hat r}}^t \widehat{MWTP} (s;F_1(s))f_1(s)ds+F_1(t_{\hat r})\hat{r} \\
  =	\int_{t_{\hat r}}^t s\Big(1+\eta \lambda F_1(s)+\eta(1-F_1(s)) \Big)f_1(s)ds + [(1+\eta)+\eta(\lambda-1) F_1(t_{\hat r})/2]F_1(t_{\hat r})t_{\hat r}, \notag
	\end{eqnarray}
for all $t\geq t_{\hat r}$.

Given these expected payments, the seller chooses the threshold type to maximize her profit:
\begin{equation}\label{upobjective}
	N \times \Big( \max_{t_{\hat r}} \int_{t_{\hat r}}^{\hat{t}}  F_1(t)\hat \beta(t) f(t)dt \Big)+  F^N(t_{\hat r})  t^S
\end{equation}

Notice that $F_1(t)\hat \beta(t)$ represents an upper bound on the expected payment of a type-$t$ bidder since $\hat \beta(t)$ is the (pseudo) bidding function of a posited equilibrium where every bidder type is exposed to the attachment effect, even those who are excluded by the reserve price. Indeed, in this hypothetical case, the threshold type is the same as in the risk-neutral benchmark since its determination is not affected by the attachment effect. Hence, the solution to the above problem provides an upper bound on the seller's profit, as formally stated in the next proposition.

\begin{proposition}\label{upper_bound_maximized}
	In the FPA, under PPE, a strict upper bound on the seller's objective function is given by the maximum value of \eqref{upobjective}.  
\end{proposition}

%We then show that the seller gains her payoff with a secret reserve price that puts arbitrarily little mass on low types.  
Next, we show that the seller can garner a profit arbitrarily close to the maximum value of \eqref{upobjective} by using secret and random reserve prices. Specifically, before buyers submit their bids, the seller publicly announces the distribution of the reserve price and commits to drawing a reserve price according to this distribution; then, after the buyers submit their bids, the seller reveals the realization of the reserve price. While such a selling mechanism undoubtedly requires some commitment on the part of the seller, there are some real-world examples where sellers seem to have such commitment power.\footnote{The commitment power needed to implement a random reserve-price regime is similar to the commitment power that the monopolist in Heidhues and  K\H{o}szegi (2014) and Rosato (2016) needs to implement a stochastic pricing strategy. Hancart (2022) shows that such commitment power is essential in their setting.} For instance, Li and Perrigne (2003) study timber auctions conducted by a French government agency who naturally has commitment power, and where the reserve price is revealed only after all bids are submitted.\footnote{Li and Perrigne (2003) report that in such auctions interested bidders would first submit sealed bids; the auctioneer would then open the bids and announce the reserve price. The government agency did not use a fixed rule to determine the reserve price but instead considered various factors, such as market conditions and its own financial constraints. See also Andreyanov and Caoui (2022) for examples of sellers committing to a distribution of secret reserve prices.} The next proposition formally states the result.

\begin{proposition}
\label{RRP} In the FPA, under PPE, there
exists a distribution of random and secret reserve prices that yields a revenue arbitrarily close to the maximum value of \eqref{upobjective}.
\end{proposition}

To gain some intuition,
consider an FPA with a secret reserve price drawn from a commonly known distribution over some interval $[\underline{r},\overline{r}]$; then, in a symmetric equilibrium in increasing strategies, for each possible realization of the reserve price, $r \in [\underline{r},\overline r]$, there exists a corresponding threshold type, $%
\tilde{t}_{r}$, whose bid coincides with $r$.\footnote{We start by positing the existence of an equilibrium such that for every $r$ there exists a $\tilde{t}_{r}$ for which $\beta(\tilde{t}_{r})=r$; then, using the implied properties of this equilibrium, we verify its existence by deriving a closed-form expression for the bidding function.}  Hence, a bidder with a type below the highest such threshold type, $\tilde{t}_{\overline r}$,  wins the good if both (i)
her type is larger than that of all other bidders, and (ii) she bids higher than
the secretly drawn reserve price.
In what follows, it will prove convenient to directly work with the implied distribution of threshold types. That is, suppose that the seller first draws $\tilde{t}_r \in [\tilde{t}_{\underline{r}},\tilde{t}_{\overline{r}}]$ according to some distribution $F_{0}$. Then, the seller computes the reserve price $r(\tilde t_{r})$ that matches the equilibrium bid of a bidder with type $\tilde t_r$, who expects to win the good with probability $q(t) =F_{1}(\tilde {t}_r)F_{0}( \tilde {t}_r)$.\footnote{The equilibrium bid depends on the distribution of the reserve prices itself; see equation \eqref{shillb} below.} %states this distribution in terms of primitives up to $F_0$.} 

%In this section we show that the seller can achieve an even higher revenue by using `uncommon selling mechanisms'. We focus on two such mechanisms. The first one is a standard auction with a secrete and random reserve price regime. The second implementation is an auction with public reserve price- followed by price-posting.

 %Recall that the attachment increases the seller's opportunity cost of raising
%the reserve price. A secrete and random reserve-price scheme allow the sellers to raise more revenue
%than under a deterministic reserve price, as the following proposition states.

Note that the above framework nests a public reserve price as a special case that occurs if all probability mass is on one threshold type, say, $\tilde t^p_r \in [\tilde{t}_{\underline{r}},\tilde{t}_{\overline{r}}]$; that is, $F_0(\tilde t_r)=0$ if $\tilde t_r<\tilde t_r^p$ and  $F_0(\tilde t_r)=1$  otherwise. However, as argued in Section~\ref{PPE}, with such a discontinuous threshold-type distribution, the seller forgoes the attachment effect of the marginal bidder with type $\tilde t^p_r$. Indeed, the implied winning probability, $q(t)$, jumps from 0 to $F_1(\tilde t_r^p)$ at the threshold type and, in a PPE, this type has the same attachment level as a type that loses the auction for sure. More generally, consider an arbitrary threshold-type distribution that might have discontinuous jump points. At each such jump point, the corresponding threshold type has the attachment level of the type immediately below her. As the seller's revenue increases in the bidders' attachment level, the seller prefers continuous distributions that smooth out jump points (see the left panel of Figure~\ref{fig:F1}).

In fact, in the proof of Proposition~\ref{RRP} we show that the seller can achieve a revenue arbitrarily close to the maximum value of \eqref{upobjective} by using a continuous distribution of threshold types. This distribution is such that every bidder type expects to win the auction with strictly positive
probability, as tiny as that might be; that is, $\tilde{t}_{\underline{r}}=\underline{t}$. More precisely, the distribution is
such that bidders with types strictly below $\tilde{t}_{\overline r}$ expect to win the auction
with a small probability, and this probability steeply increases for types in a neighborhood below $\tilde{t}_{\overline r}$. Moreover, the seller chooses $\tilde{t}_{\overline r}$
such that the virtual value of the largest threshold type is equal the seller's value; i.e., $\tilde{t}_{\overline r}=(1-F(\tilde{t}_{\overline r}))/f(\tilde{t}_{\overline r})+t^S$. One way for the seller to implement such a distribution of threshold types is to use the CDF $%
F_{0}(\tilde{t}_r)=(\frac{\tilde{t}_r}{\tilde{t}_{\overline{r}}})^{K}$, with $K\in \mathbb{R}_{+}$
and \textquotedblleft large\textquotedblright. The right panel of Figure~\ref{fig:F1} depicts $F_{0}$ when $\tilde{t}_{\overline{r}}=0.5$
and $K=30$.%\footnote{In the proof of Proposition~\ref{RRP} we characterize a strict upper bound on the seller's revenue given any reserve price, and then show that by using the distribution $F_{0}$ the seller achieves a revenue arbitrarily close to that upper bound.}

Therefore, with a secret and random reserve price, all bidders with type below the largest threshold type $\tilde{t}_{\overline{r}}$ expect to
win with strictly positive probability and are thus exposed to
(potential) losses. In particular, the steep increase of $q(t)$ from
(almost) zero to (almost) $F_{1}(\tilde{t}_{\overline{r}})$ in the neighborhood below $%
\tilde{t}_{\overline{r}}$ ensures that types slightly below $\tilde{t}_{\overline{r}}$ have an
incentive to %mimic larger types by bidding
bid aggressively in order to reduce their potential losses.%\footnote{The intuition for this result is similar to that behind the random-sales results of Heidhues and K\H{o}szegi (2014).}

 A second rationale behind the steep increase in the reserve price follows from the usual
rationing effect that is also present under risk neutrality. By imposing a larger
minimal bid, the seller increases the competitive
pressure on the bidders' side at the cost of decreasing the probability of trade. In the risk-neutral benchmark, where the attachment
effect is absent, the optimal resolution of this
trade-off entails excluding all types with virtual values below the seller's value. With the
above-described secret and random reserve price, the attachment effect does not modify
the seller's trade-off since all bidder types are exposed to it. However, in contrast to the case of a public and deterministic reserve price, the probability of receiving the good does not drop from $F_{1}(\tilde{t}_{\overline{r}})$ to 0 if the threshold type marginally lowers her bid. Thus, one might wonder whether secret and random reserve prices also intensify the competitive pressure on the bidders' side in the same way as a public reserve price. The answer is yes since, as the distribution of the secrete reserve price increases steeply below the threshold type, if a bidder with such type were to marginally lower her bid, her probability of receiving the good would suddenly drop to almost zero.  %Moreover, whether the probability of receiving
%the good(as with a public reserve price) or it increases more smoothly (as under the distribution of the secret reserve price described above) is of second-order importance for the goal of intensifying the competitive pressure on the bidders' side.
 
\smallskip
{\small 
\begin{figure}[thb]
    \centering
    \begin{subfigure}[t]{.5\textwidth}
      	\begin{tikzpicture}  [xscale=5,yscale=3]
	\node [left] at (0,.6) {$F_0(\tilde t_r)$};
	\node [below] at (1.1,-.0) {$ \tilde t_r $};

	%\fill (0,0) circle [radius=.35pt];
	%\node[left] at (.05,-.06) {$0$};
	%\node[left] at (1.05,-.06) {$1$};
	
	\draw [<->] (0,1.1) -- (0,0) -- (1 ,0);
	\draw[ dashed, black, domain=0 :1/6] plot (\x, {  .5*(6*\x)^(30) });
	\draw[ dashed, black, domain=1/6:0.5] plot (\x, { .5*(3*(\x-1/6))^30+.5 });
	\draw[ dashed,black, domain=.5:1] plot (\x, {1});
	\draw[ thick, red, domain=0 :1/6] plot (\x, { 0});
	\draw[ thick, red, domain=1/6:0.5] plot (\x, {  .5 });
	\draw[ thick,red, domain=.5:1] plot (\x, {1});
	\fill (1/6,.5) circle [radius=.35pt];
	\fill (0.5,1) circle [radius=.35pt];
	\draw [-] (1,-0.01 ) -- (1,.01) ;
	\node[below] at (1,-.01) {$ 1$};
	\draw [-] (-.01,1 ) -- (.01,1) ;
	\node[left] at (-.01,1) {$ 1$};
	\draw [ultra thick,-] (1/6,-0.02 ) -- (1/6,.02) ;
	\node[below] at (1/6,-.02) {$  \tilde{t}_{r_1}$};
	\draw [ultra thick,-] (0.5,-.02 ) -- (0.5,.02) ;
	\node[below] at (0.5,-.02) {$ \tilde{t}_{r_2}$};

\end{tikzpicture} %
     \caption{Arbitrary Distributions.}
    \end{subfigure}%
    \begin{subfigure}[t]{.5\textwidth}
         	\begin{tikzpicture}  [xscale=5,yscale=3]
	\node [left] at (0,.6) {$F_0(\tilde t_r)$};
	\node [below] at (1.1,-.0) {$ \tilde t_r $};

	%\fill (0,0) circle [radius=.35pt];
	%\node[left] at (.05,-.06) {$0$};
	%\node[left] at (1.05,-.06) {$1$};
	
	\draw [<->] (0,1.1) -- (0,0) -- (1 ,0);
	\draw[ thick, black, domain=0 :.5] plot (\x, {   (2*\x)^(30) });
	%\draw[ dashed, black, domain=1/6:0.5] plot (\x, { .5*(3*(\x-1/6))^30+.5 });
	\draw[ thick,black, domain=.5:1] plot (\x, {1});
	% \draw[ thick, red, domain=0 :.5] plot (\x, { 0});
	%\draw[ thick, red, domain=1/6:0.5] plot (\x, {  .5 });
	%\draw[ thick,red, domain=.5:1] plot (\x, {1});
	% \fill (1,.5) circle [radius=.35pt];
	%  \fill (0.5,1) circle [radius=.35pt];
	\draw [-] (1,-0.01 ) -- (1,.01) ;
	\node[below] at (1,-.01) {$ 1$};
	\draw [-] (-.01,1 ) -- (.01,1) ;
	\node[left] at (-.01,1) {$ 1$};
	\draw [ultra thick,-] (.5,-0.02 ) -- (.5,.02) ;
	\node[below] at (.5,-.02) {$\tilde{t}_{\overbar{r}}= 1/2$};

\end{tikzpicture} %
        \caption{Optimal Distribution for Unif$[0,1]$ and $t^S=0$.}
    \end{subfigure}%
  \caption{ {\footnotesize The solid red lines in the left panel depict the distribution of threshold types under an (arbitrary) secrete reserve price scheme with two reserve prices, $r_1$ and $r_2$. The dashed curve is a continuous approximation that leaves the seller with strictly larger revenue and uses infinitely many reserve prices. The right panel depicts the optimal CDF of the threshold types when bidders' types are distributed according to a Unif$%
[0,1]$. The induced distribution of secret reserve
prices, $\protect\beta_I^{\ast}(\tilde t_r)$, can be
obtained from \eqref{shillb}. }}
\small{\label{fig:F1}}
\end{figure}
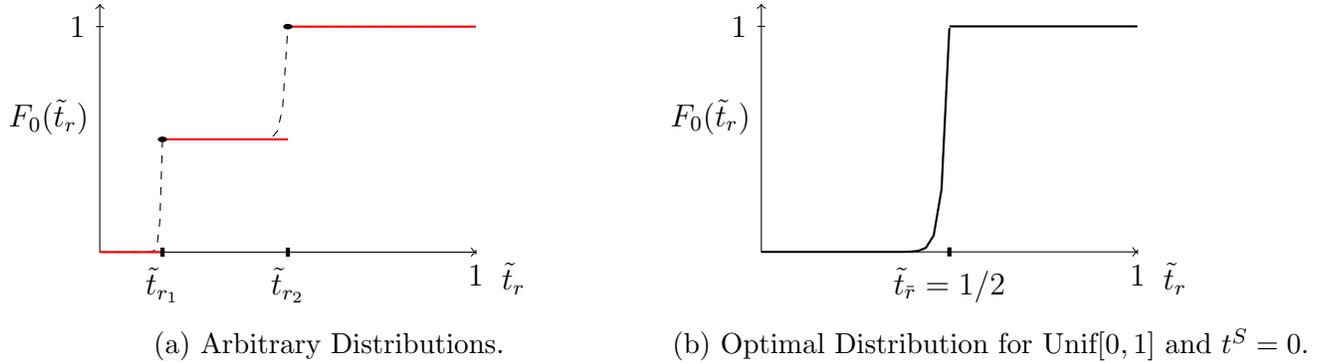}

In order to obtain the distribution of the secret reserve price, start by fixing $F_0$, the distribution of the threshold types $\tilde{t}_r$. Then, substitute the drawn $\tilde{t}_r$
into the equilibrium bidding function that applies without a reserve
price, but where a type-$t$ bidder wins the auction with probability $%
q(t)=F_{0}(t)F_{1}(t)$; that is, for $\tilde{t}_r \in \lbrack \underline{t},\tilde{t}_{\overline{r}}]$%
, we have 
\begin{equation}
\beta _{I}^{\ast }(\tilde{t}_r)=\frac{\int_{\underline{t}}^{\tilde{t}_r}[1+\eta
\lambda q(x)+\eta (1-q(x)] q^{\prime }(x)xdx}{%
q(\tilde{t}_{r})}.  \label{shillb}
\end{equation}

In equilibrium, bidders correctly anticipate that the seller implements reserve price $r=\beta_{I}^{\ast}(\tilde{t}_r)$ when drawing $\tilde t_r$ according to CDF $F_0(\tilde t_r)$, and their
optimal response is given by the bidding function in (\ref%
{shillb}), replacing $\tilde t_r$ with a buyer's type $t$.

As argued by Bajari and Horta\c{c}su (2004), secret reserve prices are common in internet auctions. However, different from our result,
these secret reserve prices are usually deterministic. Yet, from the bidders'
perspective, the secret reserve price might appear as random if they do not observe how precisely the
seller chooses
it. Indeed, as shown in Figure~1, the distribution of the secret reserve price has most of the mass on the upper bound of its
support, and arbitrarily little mass everywhere else. This characterization is consistent with the notion of bidders expecting the seller to make small ``mistakes'' when choosing the secret reserve
price. Moreover, an alternative way of implementing a secret and random reserve price is via ``shill bidding'', a prominent phenomenon in
 real-world auctions whereby a dummy buyer submits
 pre-specified bids on behalf of the seller.\footnote{While Ashenfelter (1989) provides some early examples of shill bidding in
 	art auctions, this phenomenon appears to be even more common in online auctions, where sellers can use multiple accounts to bid on their own items; see Grether et al. (2015). With risk-neutral bidders, this practice can be particularly effective in dynamic auction formats, as it enables a seller to adjust the reserve price based on information that emerges as the auction unfolds --- be it information about bidders' type distributions as in Graham et al. (1990), or information about the number of participating bidders as in Wang et al. (2001).} %,for instance, shill bidding allows the seller to learn about the distribution of buyers' values (Graham et al., 1990), or the number of participating bidders, thereby making the optimal reserve price depend on the number bidders in environments where virtual values are non-monotone (Wang et al., 2001).}

Thus, expectations-based reference points offer a novel explanation for the use of secret and random reserve prices. Rosenkranz and Schmitz (2007) provided an earlier rationale for secret (but not random) reserve prices; in their model, publicly announcing the reserve price plants a reference point in the bidders' minds, making it less attractive to win at a price higher than the reserve. While their explanation is also a reference-dependent one, it is based on loss aversion in money, whereas ours leverages the attachment effect. Secret reserve prices can also be rationalized under risk aversion and in common-value auctions
 \emph{if} the seller's value is privately known.\footnote{%
 	Li and Tan (2017) show that a seller with a privately known value may prefer a secret reserve
 	price to a public one when facing risk-averse buyers. The reason
 	is that, as the optimal reserve price depends on the seller's value, the
 	fact that the seller is privately informed makes the reserve price random from the buyers' perspective.  However, if the seller's value was commonly
 	known, as in our model, a public reserve price would then be optimal. A similar argument
 	holds in common-value auctions with risk-neutral bidders; see Vincent (1995).%
 } Furthermore, secret reserve prices can emerge with uninformed bidders who learn their value as the auction unfolds (Hossain, 2008), with an %efficiency-maximizing 
 uninformed seller who uses the information in submitted bids to decide whether to trade (Andreyanov and Caoui, 2022), or with competing sellers if
 not all buyers correctly anticipate the distribution of reserve prices across sellers (Jehiel and Lamy, 2015).

Our final result in this section compares the degree of bidder exclusion under the optimal secret and random reserve-price scheme with that under risk neutrality.

\begin{proposition}
\label{RRPexclusion}
Under the optimal secret and random reserve-price scheme, the following holds:
\begin{enumerate}
	\item If $t^S=0$, the probability
	of trade
	  is arbitrarily close to that under risk neutrality.
		%\item  converges to that
		%under risk neutrality as $N\rightarrow \infty $.
	 
	\item If $t^S>0$, the probability
	of trade
	\begin{enumerate}
		\item is strictly larger than that under risk neutrality;
		\item converges to a limit value larger than that
		under risk neutrality as $N\rightarrow \infty $.
	\end{enumerate}
\end{enumerate}  
\end{proposition}

The results of Proposition~\ref{RRPexclusion} and the intuition behind them are similar
to those in Propositions~\ref{P1} and \ref{N} under a public and deterministic reserve price. Indeed, because with secret reserve prices the attachment effect does not affect the optimal no-trade probability anymore, differences with the risk-neutral benchmark are driven solely by the fact that, in contrast to the bidders, the seller is not loss averse. Therefore, the level of bidder exclusion, and hence inefficiency, is lower with loss-averse bidders than with risk-neutral ones.
Finally, we re-emphasize that the analysis and the results in this section
carry over to the SPA.

\subsection{Auctions Followed by Negotiation}\label{sec:price positing}

 In this section, we show that the seller can achieve a larger revenue with an auction followed by a take-it-or-leave-it (TIOLI) negotiation, rather than by holding a standard auction with a (revenue-maximizing) public reserve price and committing to not selling the good if the reserve price is not met --- the latter being the optimal mechanism with risk-neutral bidders.

%Consider the following mechanism for fixed threshold type $t_r \in (0,\bar t)$: There is a public reserve price $r=t_r(1+\eta +\eta(\lambda-1)F_1(t_r)/4)$. If no bidder submits a bid above the reserve price, the auctioneer posts a price equals to $p=sup \lbrace (1+\eta)t | t<t_r \rbrace$ with probability $1/2$ at which a bidder can buy the good. If more than one bidder wants to buy at this price, the seller brakes ties arbitrarily.

Consider the following two-stage mechanism. The seller first runs a standard auction with a public reserve price $r$; the highest bidder wins the auction provided that her bid is (weakly) above the reserve price. If instead the reserve price is not met, with probability $\nu \in [0,1]$, the seller posts another price $p$ with $p<r$ at which any bidder can buy the good.\footnote{Note that the seller must commit to both $\nu$ and $p$ at the beginning of the mechanism.} If more than one bidder wants to buy the good at this price, the seller breaks ties uniformly.

%In the appendix (proof of Proposition~\ref{price posting}) we show that 
Fixing the reserve price $r$ and posted price $p$, there are two threshold types, $t_r$ and $t_p$ with $t_r>t_p$. The first one, $t_r$, is the bidder type that submits a bid exactly equal to the reserve price. The other threshold type, $t_p$, is the smallest type who is willing to buy the good at the posted price in case the reserve price is not met. In addition, let $q(t_p)$ be the (ex-ante) probability that type $t_p$ receives the good and let $\alpha \in (0,1)$ denote the solution to $\alpha F_1(t_r)=q(t_p)$; note that $q(t_p)$, and hence $\alpha$, depend on $\nu$.\footnote{Consider a bidder with type $t \in [t_p,t_r)$. If she accepts the posted price, the bidder receives the good with probability $1/\texttt{\#}$ where $\texttt{\#}$ is the number of competitors with types in $[t_p,t_r)$. Conditional on the event that no other buyer bids above the reserve (which happens with probability $F_1(t_r)$), $\texttt{\#}$ is a random variable that follows a binomial distribution with success probability $1-F(t_p)/F(t_r)$. Let $Pr(\texttt{\#};N-1)$ be the probability that exactly $\texttt{\#}$ out of $N-1$ bidders have type above $t_p$, conditional on having a type below $t_r$. Then, $q(t_p)= \nu F_1(t_r) \sum_{\texttt{\#}=0}^{N-1} Pr(\texttt{\#};N-1)/(\texttt{\#}+1)=\nu F_1(t_r) \sum_{v=0}^{N-1} {N-1\choose \texttt{\#}} (1-F(t_p)/F(t_r))^\texttt{\#} (F(t_p)/F(t_r))^{N-1-\texttt{\#}}/(\texttt{\#}+1)=\nu F_1(t_r)\frac{1-(F_1(t_p)/F_1(t_r))^N  }{ N  (1-F_1(t_p)/F_1(t_r)) }$ and thus $\alpha= \nu \frac{1-(F_1(t_p)/F_1(t_r))^N  }{ N  (1-F_1(t_p)/F_1(t_r)) }$.} 

Fixing $p$, $\alpha$, and $r$, the following two conditions must hold:
\begin{eqnarray}\label{thresholds}
(1+\eta) t_p=p ,\\
t_r \Big( (1-\alpha)(1+\eta)   + \eta (\lambda-1) (1-\alpha) \alpha F_1(t_r)       \Big)+   (1+\eta )\alpha   t_p = r.
\end{eqnarray}

The first condition, $(1+\eta)t_p=p$, is intuitive: in the PPE, the threshold type $t_p$ is the lowest type for which not buying at the posted price is not a personal equilibrium. Given the first condition, the second one pins down the threshold type $t_r$ as the highest type who, in equilibrium, is willing to buy at the posted price; that is, the type who, when expecting not to participate in the auction, is just indifferent between bidding the reserve price or bidding below the reserve price and buying the good at the posted price with probability $\alpha F_1(t_r)$. Crucially, notice that, in contrast to a situation without the possibility of buying at the posted price, when not participating in the auction the threshold type $t_r$ still expects to obtain the good with strictly positive probability, $\alpha F_1(t_r)$. In turn, participating in the auction by submitting a bid equal to the reserve price becomes more attractive for this bidder type, as it reduces her potential losses.

Given these threshold types, equilibrium behavior is then straightforward. Bidders whose type is strictly below $t_p$ abstain from the auction and do not accept the posted price. Bidders with types in $[ t_p,t_r)$ do not bid in the auction, but accept the posted price if the reserve price is not met. Finally, bidders with types weakly higher than $t_r$ participate in the auction by bidding $\beta(t)=\int_{t_r}^t (1+\eta +\eta (\lambda-1)F_1(s))sf_1(s)ds+F_1(t_r)r$. 

%In the appendix (proof of Proposition~\ref{price posting}), we show that the mechanism features the following equilibrium.
%Bidders with types weakly above $t_r$, submit the bid. Bidder types strictly below $t_r$ abstain from the auction. Bidder type $t_r^- \equiv \sup \lbrace t|t<t_r \rbrace$ accepts the price in case it is posted.

  The next proposition states that, by exposing more bidders to the attachment effect, the above described selling mechanism can generate more revenue than a standard auction with a public reserve price.

\begin{proposition}\label{price posting}
	%Consider either an FPA or an SPA. Under PPE, for every auction with a public reserve price and corresponding threshold type, say $t_r^p$, there exists an auction followed by a TIOLI that raises more revenue.
 Consider either an FPA or an SPA. Under PPE, for every auction with a public reserve price, there exists an auction followed by a TIOLI that raises more revenue.
\end{proposition}
 
 Proposition~\ref{price posting} states that there exists choices of $\nu$ and $p$ (with an implied $\alpha)$, such that the seller strictly benefits from using a TIOLI mechanism after the auction. %Indeed, as we show in the proof of Proposition~\ref{price posting}, when the seller chooses $p \rightarrow (1+\eta)t_r^p$ and $\alpha=1/2$, which is not necessarily the optimal choice, the seller's revenue is larger than under no price-posting. 
 
 %A real-world example of auctions followed by price posting if no one bids above the reserve is provided by the real estate market where often, if a property does not meet the reserve price at the auction, the seller may choose to negotiate with interested parties who did not meet the reserve. In such cases, the auction is not necessarily the final word on the sale price of the property. While the best negotiation strategy might not be feasible, our model predicts that there should be some post-auction negotiation. Moreover, if bidders are risk neutral or risk averse, a seller would optimally commit not to negotiate with bidders if no one bids above the reserve. In contrast, Proposition~\ref{price posting} shows that such negotiations are not a sign that a seller does not have the necessary commitment power; rather, such post-auction negotiations are part of an optimal mechanism if bidders are expectations-based loss averse.

There are many real-world examples of auctions followed by some form of negotiation if the reserve price is not cleared. For instance, sellers may choose to negotiate with interested parties who did not meet the reserve; see Elyakine et. al (1997) on timber sales, Bulow and Klemperer (1996) on fine-art auctions, and Ashenfelter and Genesove (1992) and Ong (2006) on real-estate auctions.\footnote{Moreover, while not exactly the same, eBay's ``Second Chance Offer", by allowing the seller to make an offer to the highest bidder below the (secret) reserve, is in a similar spirit.} Perhaps strikingly, Proposition~\ref{price posting} shows that the auctioneer benefits from engaging in some post-auction negotiation even though a TIOLI offer is not necessarily the optimal post-auction form of negotiation from the seller's perspective. Moreover, with risk-neutral bidders a seller would optimally commit to never negotiate with them after the auction if no one bids above the reserve.\footnote{There is a substantial theoretical literature focusing on (optimal) selling mechanisms for auctioneers lacking power to commit not to trade; see, for instance, McAfee and Vincent (1997) on price-posting sellers, and Skreta (2015) and Liu et. al (2015) on sellers having access to more general mechanisms. All these contributions show that the seller is hurt by being unable to commit to not selling the good if her initial mechanism does not allocate it.} In contrast, Proposition~\ref{price posting} shows that such negotiations are not necessarily a sign that a seller does not have the necessary commitment power; rather, the possibility of such post-auction negotiations can be beneficial for the seller if bidders are expectations-based loss averse.
 
The intuition for why the seller benefits from post-auction negotiation is as follows. By granting a probability of winning also to types below the margin, the seller attaches them to the good. This induces a competitive pressure, due to psychological motives, on the threshold type's bid, as lower types now have a stronger incentive to bid more in order to reduce their losses. In other words, by not excluding a set of types below the threshold type $t_r$, the seller inflates the attachment of the threshold type, who then bids more aggressively. However, there is also a cost for the seller since the larger the probability with which a type below the margin receives the good, the lower the competitive pressure, due to the material motives, on the threshold type. The seller optimally trades off these two effects --- increasing the competitive pressure on the threshold type via the attachment effect vs. decreasing it by allowing some post-auction negotiation --- by choosing a posted price $p$ that only types that are very close to $t_r$ would accept.

 \section{Conclusion\label{conclusion}}
 This paper belongs to a recent and growing literature on the market implications of expectations-based loss aversion. Indeed, over the last decade, the model of expectations-based loss aversion developed by K\H{o}szegi and Rabin (2006, 2007, 2009) has found many fruitful applications in several areas of economics, including firms' pricing and advertising strategies (Heidhues and K\H{o}szegi, 2008, 2014; Rosato, 2016; Karle and Peitz, 2014, 2017; Karle and Schumacher, 2017), incentive provision (Herweg et al., 2010; Daido and Murooka, 2016; Macera, 2018), bargaining (Rosato, 2017; Herweg et al., 2018; Benkert, 2022), labor supply (Crawford and Meng, 2011), school choice (Dreyfus et al., 2022; Meisner and von Wangenheim, 2023), asset pricing (Pagel, 2016, 2018; Meng and Weng, 2018), and life-cycle consumption (Pagel, 2017). In particular, there have been several studies on the implications of expectations-based loss aversion in auctions; see Lange and Ratan (2010), Eisenhuth (2019) and Balzer and Rosato (2021) on sealed-bid auctions, and von Wangenheim (2021), Balzer et al. (2022) and Rosato (2023) on dynamic ones.

 While the prior literature has mostly abstracted from considering reserve prices, the focus of our paper is on how expectations-based loss aversion affects the optimal reserve price, and the resulting level of bidder exclusion, in the FPA and SPA.\footnote{%
 	Three notable exceptions are Rosenkranz and Schmitz (2007), Eisenhuth (2019) and Muramoto and Togo (2022).
 	Using a different solution concept, Choice-acclimating personal equilibrium (CPE), Eisenhuth (2019) shows that the optimal reserve price in the all-pay auction
 	implies the same no-trade probability as under risk neutrality, while Muramoto and Togo (2022) focus on asymmetric auction design, allowing for bidder-specific reserve prices. In Rosenkranz and Schmitz (2007), differently from our setting, bidders are loss averse only with respect to their monetary payment and use the (public) reserve price as a reference point.} Our analysis reveals that loss aversion delivers new implications for the design of optimal auctions that are likely to be of interest for both theorists as well as practitioners. In particular, we find that random and secret reserve prices outperform deterministic and public ones in both the FPA and SPA. Indeed, by using random and secret reserve prices, the
 seller introduces a small risk that exposes all bidders to the attachment
 effect, which in turn leads them to bid more aggressively. This result establishes a tight link between the optimal level of exclusion in an auction and the optimal monopoly pricing scheme with loss-averse consumers derived by Heidhues and K\H{o}szegi (2014) and Hancart (2022), that is analogous to the well-known one under risk neutrality. More generally, our result implies that with loss-averse agents mechanisms that level the playing field by giving every player a chance to win might be better suited to generate competitive pressure than mechanisms with steeper incentives (e.g., winner-take-all contests).
 
 If the seller is forced to use a deterministic and public reserve price, we find that its optimal level depends the number of bidders in the auction and is typically lower than the optimal reserve price under risk neutrality. Moreover, we show that sellers can raise even more revenue by committing to engage in some post-auction haggling if the reserve price is not met, as this also exposes more bidders to the attachment effect; this is in stark contrast to the case of risk-neutral (or risk-averse) buyers, where post-auction negotiations are never optimal. Hence, expectations-based loss aversion rationalizes several features of reserve prices observed in real-world auctions which are hard to reconcile with the classical risk-neutral and risk-averse frameworks.

\newpage

{\normalsize 
%TCIMACRO{\TeXButton{baselineskip}{\baselineskip}}%
%BeginExpansion
\baselineskip%
%EndExpansion
=14pt }

{\normalsize \appendix
}

\section{Appendix A{\label{proofs}}}

{\small \textbf{Proof of Lemma~\ref{L1}} 
\begin{proof}
  
 Let $P(t)\equiv F_1(t)\beta_I^{\ast}(t)$ be the expected payment from a bidder with type $t \geq t_r$. We know from Balzer and Rosato (2021) that in the FPA, a bidder's first-order condition satisfies
  \begin{eqnarray}\label{D1}
  \frac{ (1+\eta \lambda F_1(t) +\eta [1-F_1(t)])f_1(t) t}{1+\eta^m \lambda^m} = P'(t)  .
  \end{eqnarray} 
  The solution to differential equation (\ref{D1}) is
  \begin{eqnarray*}
  P(t)=   \int^t_{t_r}   (1+\eta \lambda F_1(x) +\eta [1-F_1(x)])f_1(x)x dx+C ,,
  \end{eqnarray*}
  where $C$ is a constant.
  Since $P(t)=F_1(t)\beta_I^{\ast}(t)$, the constant satisfies $P(t_r)=F_1(t_r)\beta_I^{\ast}(t_r)=F_1(t_r)r$. Thus, $C=P(t_r)  $.

\end{proof}
\textbf{Proof of Proposition~\ref{P1}} 
\begin{proof}
Fix $t_r$ with corresponding $r$ (see equation \eqref{reserve}). The seller's objective is:
\begin{eqnarray}\label{R}
 N \times \Big(\int^{\bar t}_{t_r} \Big[  \int^t_{t_r} \lbrace f_1(x)x (1+F_1(x)\lambda  \eta  +[1-F_1(x)]\eta ) \rbrace   dx   +     F_1(t_r)  t_r(1+\eta )  \Big] f(t)dt + \frac{F(t_r)^N}{N}  t^S\Big)   
\end{eqnarray}
Notice first that the $t_r$ which maximizes \eqref{R} is either $t_r=\underline{t}$, $t_r=\bar t$ or an interior solution. %\footnote{$t_r=\bar t$ cannot be a solution, as it implies zero revenues.} 
In the first case, the derivative of \eqref{R} is negative at $t_r=\underline{t}$, in the second-case it is positive at $t_r=\bar t$, while in the third case the derivative is zero at an interior solution and negative for a slightly larger $t_r$. Moreover, define 
	\begin{eqnarray}\label{Rdef}
	\tilde{V}(t_r)\equiv \frac{1}{ f(t_r) } \int^{\bar t}_{t_r}  \Big(1-   \frac{\eta(\lambda-1) }{1+\eta }   f_1(t_r)t_r \Big)         f(t)dt .
	\end{eqnarray}
It is straightforward to show that the derivative of \eqref{R} with respect to $t_r$ is increasing in $t_r$ if and only if
 \begin{eqnarray}\label{ott}
 	\tilde{V}(t_r)+ \frac{t^S}{1+\eta}- t_r\geq 0.
 \end{eqnarray}

   Note that equation~\eqref{ott} rules out the potential solution $t_r=\bar t$ as $\tilde{V}(\bar t)=0$ and $\bar t>t^S$. Using expression \eqref{ott}, we see that the seller's revenue increases in $t_r$ as long as
 \begin{eqnarray}\label{t}
  V (t_r)-t^S\leq  -t^S \frac{ \eta  }{1+\eta }  - \frac{\eta (\lambda -1)}{1+\eta } \frac{[1-F(t_r)]}{f(t_r)} f_1(t_r)t_r,
 \end{eqnarray}
 where $V(t_r)=t_r-[1-F(t_r)]/f(t_r)$ is the `virtual value'.
 If $t_r=\underline{t}=0$, the left-hand side is lower than the right-hand side.
 
  The right-hand side of equation~\eqref{t} is negative. This implies that at the optimal threshold type, $t_r^{\ast}$, (no matter whether it is at the lower bound or in the interior) we have $ V(t_r^{\ast})-t^S<0$. Since $V$ is an increasing function, it must be that $t_r^{\ast}< t^{RN}$ where the optimal threshold is interior, i.e., $V(t^{RN})-t^S=0$ (because $\underline{t}=0$).
 
  Finally, notice that for $\lambda$ sufficiently large, we have that $t_r^*\rightarrow \underline{t}=0$ as, for any $t>0$, the right-hand side of \eqref{t} becomes arbitrarily negative when $\lambda$ becoming arbitrarily large. Obviously, for such sufficiently large $\lambda $, the optimal reserve price, $r^{\ast}$, is $r^*=(1+\eta)t_r^*<t^S\leq r_{RN}$.

  \end{proof}}

\textbf{Proof of Proposition~\ref{N}} 
\begin{proof}

If, for some $N$, the optimal $t_r^{\ast}=0$ it trivially follows that it is weakly increasing in the number of bidders.
Thus, without of loss of generality, that the optimal threshold type is in the interior. We know that \eqref{t} from the proof of Proposition~1 holds with equality, so that 
\begin{eqnarray} \label{interior}
	V (t_r^*) -t^S \frac{ 1}{1+\eta }  + \frac{\eta (\lambda -1)}{1+\eta } \frac{[1-F(t_r^*)]}{f(t_r^*)} f_1(t_r^*)t_r^*=0,
\end{eqnarray}
Moreover, note that the derivative of the left-hand side w.r.t. $t_r$, say $LHS_{t_r}(t_r^*)$, is positive at the optimal $t^*_r$, as otherwise, increasing $t_r$ at $t^*_r$ increased the seller's profit. 
Applying the implicit function theorem to \eqref{interior} we have
\begin{eqnarray}
    \frac{dt_r^*}{dN}= -\frac{LHS_N(t_r^*,N)}{LHS_{t_r}(t_r^*)},
\end{eqnarray}
where the term $LHS_N(t_r^*,N)$ denotes the derivative of the left-hand side of \eqref{interior} w.r.t. $N$ at $t_r=t_r^*$. We thus need to a derive a condition ensuring that this term is negative. Note that $LHS_N(t_r^*,N)=\frac{\eta(\lambda-1)}{1+\eta} \frac{[1-F(t_r^*]}{f(t_r^*)} t_r^*\frac{df_1(t_r^*)}{dN} $. We thus need to figure out when $\frac{df_1(x)}{dN} =\frac{d ((N-1)F^{N-2}(t_r^*)f(t_r^*))}{dN}<0$, or, equivalently, when $\frac{dln(f_1(t_r^*))}{dN}=1/(N-1) +ln(F(t_r^*))<0 $, which is the condition from the proposition.

 % If $\eta^m=0$, then equation \eqref{t} from the proof of Proposition~\ref{P1} applies. Thus, at the optimal threshold type, $t_r^{\ast}$, we have $V(t_r^{\ast})-t^S<0$. Since $V$ is an increasing function, less types than under risk neutrality are excluded. Moreover, note that the right-hand side of \eqref{t} depends on $f_1(t_r)$ which itself depends on $N$. Note that, for any $t_r<\bar t$, $f_1(t_r) \rightarrow 0$ if $N \rightarrow \infty$. Thus, using \eqref{t}, we have that $t_r^* \rightarrow t^{RN}$ with $N \rightarrow \infty$ if and only if $t^S=0$.
  
 %In addition, using \eqref{t} it is straightforward to see that if $N \rightarrow \infty$, and thus $f_1(t_r)\rightarrow 0$, the optimal $t_r^{\ast}$ is in the interior. Using \eqref{rt} we see that, if $N\rightarrow \infty$, the optimal reserve price, $r^{\ast}$, satisfies $r^{\ast}\rightarrow t^S+(1+\eta^g)\frac{1-F(t^{\ast}_r)}{f(t^{\ast}_r)}\geq t^S+(1+\eta^g)\frac{1-F(t^{RN})}{f(t^{RN})}>r_{RN}$, as $t^{\ast}_r<t^{RN}$ and $[1-F(t)]/f(t)$ is decreasing in $t$.

  \end{proof}

{\small \textbf{Proof of Proposition~\ref{upper_bound_maximized}  } }

\begin{proof}
 We prove the proposition in two steps. First, we present a function that bounds the seller's expected-revenue function (with the reserve price being the argument) from above. Second, we find the public reserve price that maximizes that upper bound.   
 
 \textbf{Step 1: Bounding the Seller's Revenue}.
 Recall that the seller's revenue under a deterministic reserve price is given by
 \begin{eqnarray*} 
 	\int^{\bar t}_{t_r} \Big( \int^t_{t_r} \lbrace F_1'(s)s(1+F_1(s)\lambda  \eta  +[1-F_1(s)]\eta )\rbrace  ds+     F_1(t_r)  t_r(1+\eta ) \Big) f(t)dt.
 \end{eqnarray*}
 
 We replace $F_1(t_r)  t_r(1+\eta )$ with an upper bound, say, $\hat P(t_{\hat r})$, where
 \begin{eqnarray}\label{bound}
 \hat P(t_{\hat r})=   (1+\eta ) t_{\hat r}F_1(t_{\hat r})+ \eta(\lambda-1)  t_{\hat r} F_1(t_{\hat r})^2/2.
 \end{eqnarray}
  Further define $h(s) \equiv F_1'(s)sv(s)$ with $v(s) \equiv 1+\eta +\eta(\lambda-1)  F_1(s)$, then the expected revenue's upper bound, $\hat R(t_{\hat r})$, (divided by $N$) is

 \begin{eqnarray}\label{RB}
 	 \hat R(t_{\hat r})= \int_{t_{\hat r}}^{\bar t} \Big(     \int^t_{t_{\hat r}} h(s)   ds +    \hat P(t_{\hat r})   \Big) f(t)dt .
 \end{eqnarray}
 
 \textbf{Step 2: Maximizing the upper Bound}.
 
 The derivative of $ t^S F(t_{\hat r})^N/N+  \hat R(t_{\hat r})  $ with respect to $t_{\hat r}$ is
 \small{ \begin{eqnarray}\label{op}
 		t^Sf(t_{\hat r}) F_1(t_{\hat r}) -f(t_{\hat r})  \hat P(t_{\hat r})  +    \Big( -h(t_{\hat r})   +  \hat P'(t_{\hat r})   \Big) (1-F(t_{\hat r}))\notag \\
 		= t^Sf(t_{\hat r}) F_1(t_{\hat r}) -f(t_{\hat r})c \hat P(t_{\hat r})+       \frac{  \hat P(t_{\hat r})}{t_{\hat r}} (1-F(t_{\hat r})) \notag \\
 		= t^Sf(t_{\hat r}) F_1(t_{\hat r})  -    \frac{\hat P(t_{\hat r})}{t_{\hat r}}\Big( f(t_{\hat r})t_{\hat r}-[1-F(t_{\hat r})] \Big) ,
 \end{eqnarray}}
 where we used that 
 \begin{eqnarray*}
 	\hat P'(t_{\hat r})
 	= \frac{ \hat P(t_{\hat r})}{t_{\hat r}} +  (1+\eta)    F_1'(t_{\hat r}) t_{\hat r}+\eta(\lambda-1)F_1(t_{\hat r})f_1(t_{\hat r})t_{\hat r} 
 \end{eqnarray*}
 
 Recall that $V(t)= t-[ 1-F(t)]/f(t)$. \eqref{op} reveals that the optimal threshold type of the upper bound, say $t_{\hat r}^{\ast}$, satisfies
 \begin{eqnarray}\label{p5}
 	V( t_{\hat r}^{\ast})=    \frac{F_1(t_{\hat r}^{\ast})t_{\hat r}^{\ast}}{ \hat P(t_{\hat r}^{\ast}) }t^S\ .
 \end{eqnarray}
\end{proof}
 
{\small \textbf{Proof of Proposition~\ref{RRP}  } }

{\small 
\begin{proof}

 We show that the seller can achieve the maximized upper bound with the following secret and random reserve-price scheme.
For given $K \in \mathbb{R}^+$, introduce a random variable, $\tilde{T}_r$, with realization $\tilde{t}_r \in [\underline{t},\tilde{t}_{\overline{r}}]$, with $\tilde{t}_{\overline{r}}=t_{\hat r}^{\ast}$ (as defined in the proof of Proposition~\ref{upper_bound_maximized}), drawn according to CDF $F_0(\tilde{t}_r )=(\frac{\tilde{t}_r }{\tilde{t}_{\overline{r}} })^K$. Moreover, for $q(t)=F_0(t)F_1(t)$ consider 
\begin{equation} \label{bid}
 \beta_I^{\ast}(\tilde{t}_r )=\Big(\int^{\tilde{t}_r }_{\underline{t}} ( 1+\eta  \lambda  q(s)+ \eta [ 1-q(s)] )  q'(s) s ds +  \Big)/ q(\tilde{t}_r).
 \end{equation}
 If the secret reserve price follows the random function $\beta_I^{\ast} \circ \tilde{T}_r: [\underline{t},\tilde{t}_{\overline{r}}] \mapsto \Delta([\beta_I^{\ast}(\underline{t}),\beta_I^{\ast}(\tilde{t}_{\overline{r}})])$, then, in a symmetric equilibrium in increasing strategies, type-$t$ bidder expects to win the auction with probability $q(t)=F_0(t)F_1(t)$, which by Lemma~1 reinforces \eqref{bid} as the equilibrium bidding strategy (when replacing $F_1(t)$ with $q(t)$ and replacing the support of \eqref{bid} with $[\underline{t},\bar{t}]$). Define $h_q(s)\equiv q'(s)sv(s)$ with $v_q(s) \equiv 1+\eta+\eta(\lambda-1) q(s)$. Then, replacing $F_1$ with $q$ it is straightforward to observe that the expected payment, $P(t)$, satisfies 
\begin{equation} \label{10}
  P(t)=   \int^t_{\underline{t}} h_q(s)  ds    + P(\underline{t}) ,
\end{equation} 
for some constant $P(\underline{t})$ (see Lemma~\eqref{L1}).
Obviously, $P(\underline{t}) =\beta_I^{\ast}(\underline{t}) q(\underline{t})=0$, as $q(\underline{t})=0$. %where  $\beta_I^{\ast}(\underline{t})=\frac{1+\eta^g}{1+\eta^m \lambda^m} \underline{t}$.  
%Moreover, for given $K$ and $P(t_r )$, choose $\bar r$ such that $q(t_r) \bar r= P(t_r)$, i.e., $\bar r= P(t_r)/q(t_r)$. It then follows that for any $t<t_r$ the threshold reserve is $r(t)=P(t)/q(t)$ lies in $[\underline{r},\bar r]$.

Finally, we want to show that for $K \rightarrow \infty$ the seller's payoff converges to the maximized upper bound \eqref{RB}. Indeed, first note that

\begin{eqnarray}\label{secretpay}
  P(t)=     \int^t_{\tilde t_{\bar r} } h_q(s) ds  +  \underbrace{\int^{\tilde t_{\bar r} }_{\underline{t}} h_q(s) ds}_{  P(\tilde t_{\bar r} ) }      .
\end{eqnarray}

Observe first that, for any $t\geq \tilde t_{\bar r}=t_{\hat r}^{\ast}$, we have $q(t)=F_1(t)$, implying $h_q(t)= h(t)$. Thus, the expectation of the first integral of $  P(t)$ stated in \eqref{secretpay} is equal to $ \hat R(t_{\hat r}^{\ast})- \hat P(t_{\hat r}^{\ast})  (1-F(t_{\hat r}^{\ast}))$ (see \eqref{RB}).

It thus remains to show that  $  P(\tilde t_{\bar r} ) $ (stated in the under-bracket of \eqref{secretpay}) converges to $ \hat P(t_{\hat r}^{\ast})$ with $K \rightarrow \infty$. Applying partial integration reveals that

  \begin{eqnarray*}
   P(\tilde t_{\bar r})=   
(1+\eta)q(\tilde t_{\bar r} )\tilde t_{\bar r}+\eta(\lambda-1)( \frac{q(\tilde t_{\bar r})^2}{2}\tilde t_{\bar r} -\int_{\underline{t}}^{\tilde t_{\bar r} } \frac{q(s)^2}{2}ds).
  \end{eqnarray*}
  We now use that $q(t)\rightarrow 0$ if $t<\tilde t_{\bar r}=t_{\hat r}^{\ast}$.  
 Thus $   P(\tilde t_{\bar r}) \rightarrow  \hat P (t_{\hat r}^{\ast})$ and the claim follows.
\end{proof}

{\small \textbf{Proof of Proposition~\ref%
		{RRPexclusion}} }
\begin{proof}
	Consider equation~\eqref{p5} from Step 2 of Proposition~\ref{upper_bound_maximized}'s proof.
Assume that $t^S=0$. The largest optimal threshold $t_{\hat r}^{\ast}$, satisfies $V(t_{\hat r}^{\ast}) =0$, and thus the threshold coincides with the risk-neutral one.  
Now assume that $t^S>0$. The optimal threshold then satisfies $V(t_{\hat r}^{\ast})=t^S/\Big(1+\eta+F_1(t_{\hat r}^{\ast})\eta(\lambda-1)/2 \Big)<t^S$. Since $V$ is increasing in $t$, $t_{\hat r}^{\ast}$ is strictly smaller than the risk-neutral threshold type. Moreover, it is easy to observe that $V(t_{\hat r}^{\ast})$ converges to $t^S/(1+\eta)$ if $N \rightarrow \infty$ and thus is lower than the risk-neutral optimal threshold.  
\end{proof}

 {\small \textbf{Proof of Proposition~\ref%
 		{price posting}} }
 
 \begin{proof}
 	 Take any auction with public reserve price, where the seller commits not to sell the good in case no bidder meets the reserve. The corresponding threshold type is $t_r$. In the following, we show that there exists an auction followed by TIOLI negotiations that yields the seller larger revenues. For fixed $t_r$ (as in the auction without TIOLI negotiations) and $t_p$, this latter mechanism is characterized by an interval from $[t_p,t_r)$. Types in that interval receive the good with constant probability $ \alpha F_1(t_r)$, where $\alpha<1$ and pay price $(1+\eta)t_p$ if they get the good. In the following we derive the bid function of $t_r$, that is, the reserve price, and show that there exist feasible choices of $t_p$ (i.e., $p$) and $\mu$ such that the seller raises more revenue than without having TIOLI negotiations. %Note that when $\underline{t}_r \rightarrow t_r$, $\underline{q}= \alpha F_1(t_r) $ is feasible for any $\alpha<1$.
 	
 	%We first focus on the bid function up to $t_r$. Using the formula, we have
 %	$\underline{q}\beta(t)= \alpha F_1(t_r) (1+\eta^g)\underline{t}_r$ for any $t \in [\underline{t}_r,t_r]$.
 	 Note that $F_1(t_r)r  $ is determined by $\inf \lbrace t | t<t_r \rbrace$'s incentive constraints. We have
 	\begin{eqnarray*}
 		F_1(t_r)t_r-\eta \lambda  \alpha F_1(t_r) (1-  F_1(t_r))t_r+\eta F_1(t_r)(1- \alpha F_1(t_r))t_r- F_1(t_r)r \\
   \leq \alpha F_1(t_r) t_r  -  \eta \lambda \alpha F_1(t_r)(1- \alpha F_1(t_r) )t_r+\eta  \alpha F_1(t_r)(1-\alpha F_1(t_r))t_r- \alpha F_1(t_r)(1+\eta )t_p \\
 	\Leftrightarrow	(1-\alpha)F_1(t_r)t_r + \eta \lambda (1-\alpha) \alpha F_1(t_r)^2t_r + \eta (1-\alpha F_1(t_r))(1-\alpha)F_1(t_r)t_r+   (1+\eta )\alpha F_1(t_r) t_p \leq  F_1(t_r) r
 	\end{eqnarray*}
 	Dividing by $F_1(t_r)$ we have
\begin{eqnarray*}
    	t_r \Big( (1-\alpha)   + \eta \lambda (1-\alpha) \alpha F_1(t_r)   + \eta (1-\alpha F_1(t_r))(1-\alpha)   \Big)+   (1+\eta )\alpha   t_p \leq    r \\
     \Leftrightarrow t_r \Big( (1-\alpha)(1+\eta)   + \eta (\lambda-1) (1-\alpha) \alpha F_1(t_r)       \Big)+   (1+\eta )\alpha   t_p \leq    r
\end{eqnarray*}
In equilibrium, the above condition holds with equality and pins down the reserve price.
  Suppose the seller chooses $t_p \rightarrow t_r$ and $\alpha=1/2$ (as $\alpha$ continuously increases from 0 to 1 with $\mu$, there exist a feasible choice of $\mu$ to implementing $\alpha=1/2$), in which case the reserve price becomes $t_r \Big(  1+\eta    + \eta (\lambda-1)   F_1(t_r) /4      \Big)$ which is strictly larger than the counterpart under no commitment $t_r \Big(  1+\eta        \Big)$.

 	%Finally, note that by continuity of the revenue in $\underline{q}$ and $\underline{t}_r$, there exists $\underline{t}_r<t_r$ together with a feasible $\underline{q}<F_1(\underline{t}_r)+\frac{F_1(t_r)-F_1(\underline{t}_r)}{N} $ such that an auction followed by price posting leads to a strictly larger revenue an auction with an optimally chosen public reserve price.  
 \end{proof}

\end{document}